\documentclass[aps,prd,showpacs,amssymb,floatfix,nofootinbib]{revtex4}
\usepackage{graphicx,amsmath,subfigure}
\usepackage{appendix}

\newcommand{\eeq}{\end{equation}}
\newcommand{\bea}{\begin{eqnarray}}
\def\ltsima{$\; \buildrel < \over \sim \;$}
\def\simlt{\lower.5ex\hbox{\ltsima}}
\def\gtsima{$\; \buildrel > \over \sim \;$}
\def\simgt{\lower.5ex\hbox{\gtsima}}

\def\lesssim{\mathrel{\hbox{\rlap{\hbox{\lower4pt\hbox{$\sim$}}}\hbox{$<$}}}}
\def\gtrsim{\mathrel{\hbox{\rlap{\hbox{\lower4pt\hbox{$\sim$}}}\hbox{$>$}}}}
\def\alt{\mathrel{\hbox{\rlap{\hbox{\lower4pt\hbox{$\sim$}}}\hbox{$<$}}}}
\def\agt{\mathrel{\hbox{\rlap{\hbox{\lower4pt\hbox{$\sim$}}}\hbox{$>$}}}}

\def\gta{\ifmmode {\mathbin{\lower 3pt\hbox   
    {$\,\rlap{\raise 5pt\hbox{$\char'076$}}\mathchar"7218\,$}}}
    \else {${\mathbin{\lower 3pt\hbox
    {$\rlap{\raise 5pt\hbox{$\char'076$}}\mathchar"7218\,$}}}
    $}\fi}
\def\lta{\ifmmode {\,\mathbin{\lower 3pt\hbox   
    {$\,\rlap{\raise 5pt\hbox{$\char'074$}}\mathchar"7218\,$}}}
    \else {${\mathbin{\lower 3pt\hbox
    {$\rlap{\raise 5pt\hbox{$\char'074$}}\mathchar"7218\,$}}}
    $}\fi}

\begin{document}
\title{ Intermediate--mass--ratio--inspirals in the Einstein Telescope.\\ II. Parameter estimation errors.}
\author{E. A. Huerta and Jonathan R. Gair}

\affiliation{Institute of Astronomy, Madingley Road, CB3 0HA Cambridge, UK}

\email{eah41@ast.cam.ac.uk}
\email{jgair@ast.cam.ac.uk}


\date{\today}

\begin{abstract}
We explore the precision with which the Einstein Telescope (ET) will be able to measure the parameters of intermediate--mass--ratio inspirals (IMRIs), i.e., the inspirals of stellar mass compact objects into intermediate-mass black holes (IMBHs). We calculate the parameter estimation errors using the Fisher Matrix formalism and present results of Monte Carlo simulations of these errors over choices for the extrinsic parameters of the source. These results are obtained using two different models for the gravitational waveform which were introduced in paper I of this series.  These two waveform models include the inspiral, merger and ringdown phases in a consistent way. One of the models, based on the transition scheme of Ori \& Thorne \cite{amos}, is valid for IMBHs of arbitrary spin, whereas the second model, based on the Effective One Body (EOB) approach, has been developed to cross--check our results in the non-spinning limit. In paper~I of this series, we demonstrated the excellent agreement in both phase and amplitude between these two models for non--spinning black holes, and that their predictions for signal--to--noise ratios (SNRs) are consistent to within ten percent.  We now use these waveform models to estimate parameter estimation errors for  binary systems with masses $1.4M_\odot$+$100M_{\odot}$, $10M_\odot$+$100M_{\odot}$, $1.4M_\odot$+$500M_{\odot}$ and $10M_\odot$+$500M_{\odot}$ and various choices for the spin of the central IMBH.   Assuming a detector network of three ETs, the analysis shows that for a \(10 M_{\odot}\) compact object (CO) inspiralling into a \( 100M_{\odot}\) IMBH with spin \(q=0.3\), detected with an SNR of 30, we should be able to  determine the CO and IMBH masses, and the IMBH spin magnitude to fractional accuracies of \(\sim10^{-3}, \, \sim10^{-3.5}\) and  \( \sim10^{-3}\), respectively. We also expect to determine the location of the source in the sky and the luminosity distance to within \(\sim 0.003\) steradians  and \(\sim 10\%\), respectively. We also compute results for several different possible configurations of the 
detector network to assess how the precision of parameter determination depends on the network configuration.
\end{abstract}
 

\maketitle

\section{Introduction}    
Significant progress has been made over the last few years in the quest for the first direct detection of gravitational waves (GWs). During the most recent published data--taking run of the ground-based GW detectors five instruments were operational. The three LIGO detectors started their Science Run 5 (S5) in November 2005 \cite{ligo}. Subsequently, the GEO 600 detector \cite{geo} joined the S5 run in January 2006 and the Virgo detector \cite{virgo} started its Science Run 1 (VSR1) in May 2007. All five detectors worked in triple--coincidence until the beginning of October 2007. The data collected during the rest of the S5/VSR1 run, referred to as ``S5y2/VSR1'', was the first long--term observation with the worldwide network of interferometric detectors. Towards the end of 2010, the LIGO and Virgo detectors were taken offline and the upgrade to the Advanced detectors began. If the design sensitivities are achieved, Advanced LIGO and Virgo will increase the detection range of future searches by as much as a factor of 10 over those reached in S5y2/VSR1, so that the monitored volume of the universe will increase by a factor of  \(\sim 1000\). At this level of sensitivity, the detection of GW signals from binary mergers should become routine.


Seismic noise will still limit the sensitivity of the advanced detector network LIGO--Virgo--GEO to frequencies \(\gtrsim 10\) Hz \cite{AD,noise}.  Hence, upgraded ground--based advanced detectors will not be very sensitive to signals from coalescing binaries in the \(\sim 10 M_{\odot}-1000 M_{\odot}\) range, which generate GWs predominantly in the 0.1--10 Hz frequency band. These events will also be inaccessible to the space-based GW detector LISA, as this instrument will be most sensitive to events with total mass \(\gtrsim 1000 M_{\odot}\), i.e., signals in the \(0.1 \rm mHz-0.1 \rm Hz\) band. Studying GW sources in 0.1--10 Hz frequency band is needed to address an important, yet elusive, problem in astronomy, that of the existence of black holes of intermediate mass (IMBHs). There has been a long standing suspicion that these objects may form in the centres of dense stellar clusters \cite{wyller,bahcall, frank}. Some evidence supporting this possibility has come to light over the last decade, mainly from X--ray and optical observations~\cite{evidence}. Detection of these IMBHs, particularly in dense stellar clusters, may be important as they would have a direct impact in the dynamical evolution of the host cluster and lead to the generation of GWs.

There are two types of data that suggest the existence of IMBHs: i) ultraluminous X-ray sources have been observed in galaxies that are not associated with AGN, but which have fluxes which would imply accretion rates many times the Eddington limit, if the accreting object had a mass $M \lesssim 20M_{\odot}$; ii) the stellar kinematics in the centres of several globular clusters, e.g., M15 in the Milky Way, and G1 in M31, show evidence for an excess of dark mass in their centres. There are two distinct channels through which IMBHs might have formed. Firstly, they could form in the early universe through the collapse of massive, low--metallicity ``Population III'' stars. In the cold dark matter framework, galaxies formed through a series of mergers between initially low-mass objects that condensed at early cosmic times. The black holes seen today similarly grew through a combination of mergers and accretion, from initial seeds. In the Pop-III star model, the initial seed black holes are of intermediate mass, \( 100M_{\odot} - 1000 M_{\odot}\), but the initial seeds could also have been `heavy', $\sim10^5M_{\odot}$. The discovery of IMBHs present at early cosmic times would be a strong discriminator between these alternative models. The second proposed channel for IMBH formation is the runaway collision of stars in the centre of a dense stellar cluster \cite{etgair,lrs}. If IMBHs are found to be common in globular clusters, this would be an important probe of the processes taking place in such dense stellar systems. Observations in the electromagnetic spectrum will improve in coming years, but it is likely that the first robust mass determination, and hence the first convincing proof of the existence of IMBHs, will come from GW observations~\cite{evidence}.

In order to detect GWs from systems containing IMBHs, we require a detector operating in the $1-10$Hz frequency range. A design study is currently ongoing within Europe for a third-generation ground-based GW detector, the Einstein Telescope (ET) \cite{Freise:2009,Hild:2009}, to follow on from the Advanced detectors. The target is a sensitivity ten times better than that of the Advanced interferometers, and a frequency sensitivity band that stretches down into the $1-10$Hz range. The ET  will be an outstanding tool to address problems in fundamental physics, cosmology and astrophysics. It will have the capability to do the same type of science as the Advanced detectors, but with better sensitivity, and to do new science through the detection of sources at lower frequencies. The IMBH sources fall into this second category, and in this paper we will consider GWs generated from intermediate-mass-ratio inspirals (IMRIs), which are the mergers of IMBHs with lower mass compact objects, namely neutron stars or stellar--mass black holes.

For ET to detect GWs generated during IMRIs, we need accurate waveform models in the intermediate-mass-ratio regime, which are not yet available. Source modelling is an ongoing effort that has reached a fairly advanced stage for comparable-mass-ratio and extreme-mass-ratio systems. In the comparable-mass regime post--Newtonian (PN)  theory has provided waveform templates for the inspiral phase evolution, while numerical relativity (NR) can now be used to model the final few cycles of inspiral and the plunge and merger \cite{rev}. At the other extreme we have extreme--mass--ratio inspirals (EMRIs) with mass ratios $\sim10^{-6}$--$10^{-4}$. In this regime, the smaller object may complete several hundred thousand orbits at a velocity that is a significant fraction of the speed of light before crossing the horizon of the central massive black hole. Modelling the gravitational radiation of EMRIs using NR is presently impractical due to the excessive computational cost of simulating so many orbits. However, we can model these systems accurately using black hole perturbation theory, treating the mass--ratio as a small expansion parameter \cite{LRP,SFB,kludge,cons}. The sources we shall now examine lie somewhere between these two regimes.  In paper I of this series~\cite{firstpaper} we developed two waveform models that include the inspiral, merger and ringdown phases using the best of what is currently available. In the first model, valid for  CO inspirals into IMBHs of arbitrary spin, we used  the transition scheme developed by Ori \& Thorne \cite{amos}. In order to build up confidence in this waveform model, we also used the effective--one--body (EOB) approach~\cite{fai,prcs} to construct a different waveform model for non--spinning systems. The two models showed an excellent level of agreement in both the phase and the amplitude. We used these models to compute signal--to--noise ratios (SNRs) for a selection of IMRI systems. We compared results from the two models for non--spinning systems and found that the predictions were consistent to better than ten percent. We used these SNRs to estimate the number of events that could be detected by ET and found that this could be as many as hundreds of IMRI coalescences per year up to redshift \(z \sim 5\), if the systems are predominantly light and with high spin, (i.e., an IMRI with redshifted masses,  $[10+100] M_{\odot}$, and with IMBH spin parameter \(q=0.9\)), or as many as tens of IMRI coalescences at redshift \(z \sim 0.3 \), if the systems are predominantly heavy and with low spin ($[1.4+500] M_{\odot}$, with \(q=0\))~\cite{firstpaper}. In this paper, we shall address a complementary problem, namely, to calculate the precision with which an ET network will be able to estimate parameters of any IMRI sources that are detected. This is an important problem because an accurate determination of the masses of a merging binary and a measurement of the luminosity distance at which the merger is taking place will be useful for extracting science from the observations.  Since the binary systems are typically short--lived, a single ET will not be able to localise a source on the sky, so we will assume the existence of a detector network. We will consider several different network configurations in order to assess how this affects the science, which will be useful input for future design decisions.

This paper is organized as follows.  In Section~\ref{s2.1} we outline the assumptions we shall adopt for the detector and detector networks, and describe the binary systems to be used in the subsequent analysis. In Section~\ref{mod} we  present a brief description of the waveform models, introduced in paper I, that will be used again here to estimate parameter estimation accuracies. In Section~\ref{s2.2} we summarise some basics of signal analysis that will be  relevant for our studies and briefly describe the noise model we shall use for our studies, ``ET B'' \cite{etb}. In Section~\ref{s2.3} we will use our waveform models, as described in paper I~\cite{firstpaper}, to estimate the parameter errors that would arise due to detector noise, using the Fisher Matrix formalism. We present results for a 3--ET network for 12 sample IMRI systems and use four of these to explore how the results change under alternative configurations of the network. Finally, we summarize our work and discuss the implications of these results in Section~\ref{s2.4}.

\section{Assumptions}
\label{s2.1}
\subsection{Einstein Telescope Design}
\label{s2.1a}
\begin{figure*}[!htp]
\centerline{
\includegraphics[width=0.5\textwidth, angle=0,clip]{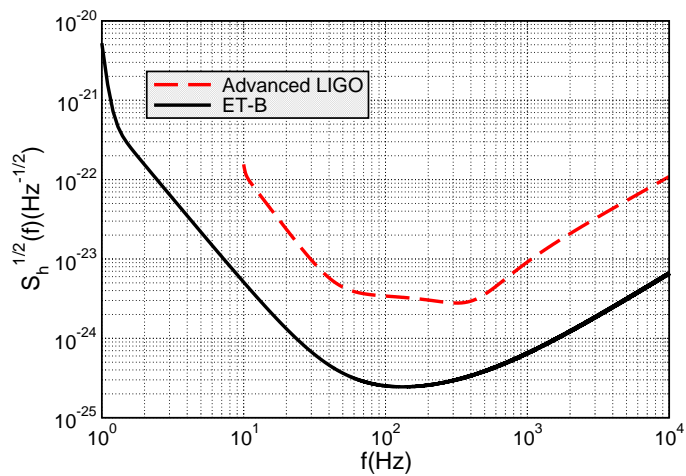}
}
\caption{Sensitivity curve for the Einstein Telescope. The Advanced LIGO noise curve is also shown for reference.}
\label{sensi}
\end{figure*}

We follow the same assumptions about the ET sensitivity that we made in paper I. The design sensitivity is usually quoted for a single right-angle interferometer with 10km arms. In this paper we shall use the  ``ET B'' sensitivity curve, as described in~\cite{etb}, which was the most up-to-date sensitivity curve available when this work started. The corresponding amplitude spectrum is plotted in Figure~\ref{sensi}. 


It is expected that ET will operate with an improved sensitivity over advanced detectors in the frequency range 1--10Hz.  Assuming a low frequency cut--off of 1Hz might be too optimistic as it is a significant technological challenge to push the sensitivity toward lower frequencies. ET may therefore only achieve sensitivity down to \(\sim 3\)Hz or higher. Given this uncertainty, we adopt a conservative approach and use a low frequency cut--off of 5Hz.

We take the response of a ``single ET'' to be that of two right-angle interferometers, coplanar and colocated but offset from one another by 45$^{\circ}$. The currently favoured ET design is for a triangular configuration comprising three 10km detectors with 60$^{\circ}$ opening angles in a single facility, as this has lower infrastructure costs \cite{Freise:2009,Hild:2009}. The response of the triangular configuration contains the same information as that of the two right-angle interferometers (assuming uncorrelated noise), but its sensitivity is a factor of $3/(2\sqrt{2}) \approx1.06$ higher. As in paper I, we ignore this factor since it is small compared to other uncertainties in the design.

IMRI signals are short--lived, so to enable extrinsic parameter estimation we need to assume the existence of a network of detectors. Our optimistic reference case will be a network of three detectors, each with the sensitivity of a single ET. We assume the detectors are sited at the geographic locations of  Virgo, LIGO Livingston and in Perth (Australia). This configuration is highly optimistic, so we shall also explore how parameter estimation accuracies are modified for four additional network configurations,  C1-C4.  These configurations are, C1: one ET at the geographic location of Virgo; C2: as configuration C1 plus a right--angle detector at the location of LIGO Livingston;  C3: as configuration C1 plus another ET at the location of LIGO Livingston;  and C4: as configuration C2 plus another right--angle detector in Perth. We will denote the reference 3-ET network as configuration C5.

\subsection{Sample IMRI systems}
\label{s2.1b}
We will present parameter estimation errors for the same twelve binary systems that we used in paper I~\cite{firstpaper}. These correspond to all possible combinations between four sets of component masses --- $1.4M_{\odot} + 100M_{\odot}$,  $1.4M_{\odot} + 500M_{\odot}$, $10M_{\odot}+100M_{\odot}$, and $10M_{\odot}+500M_{\odot}$ --- and three different values for the spin of the central IMBH --- \(q=0,\,0.3,\,0.9\). The mass and spin distributions of IMBHs are extremely uncertain, so we choose these systems to cover the range of possibilities. The small object masses are chosen to represent mergers with black holes ($10M_{\odot}$) or neutron stars ($1.4M_{\odot}$).

As discussed in paper I~\cite{firstpaper}, if IMBHs gain mass through a series of minor mergers with smaller objects, then their angular momenta will undergo a damped random walk. Under such an evolution, it is expected that IMBHs in the mass range \(100M_{\odot}-1000  M_{\odot}\) would end up with moderate spin parameters, \(q\sim 0.3\)~\cite{mandel}. We will regard this as the `typical' case and will use only that spin value when we explore how the parameter estimation accuracies depend on the network configuration. For the other spin parameters, we shall only quote results for the optimistic 3 ET network and these should be regarded as `best-case' values for the accuracy with which we may be able to determine the parameters.  

\section{IMRI waveform modelling}
\label{mod}
We will use the two waveform models developed in paper I of this series to explore the precision with which the various ET network configurations will be able to determine the source parameters. Full details of the waveforms are given in~\cite{firstpaper}, but we briefly summarise the models here. 

Both waveform models include inspiral, merger and ringdown in a consistent way.  We model the inspiral phase evolution, in both models, using the ``numerical kludge'' waveform model described in~\cite{kludge}. The ``numerical kludge'' approach consists of computing the orbital trajectory that the inspiralling body follows in the Boyer--Lindquist coordinates of the Kerr spacetime of the central black hole. We assume an adiabatic evolution in which the inspiralling body is moving on a geodesic and use the fluxes of energy \(E\) and angular momentum \(L_z\) derived in \cite{improved} to evolve the parameters of the geodesic. In this paper, we restrict our attention to circular and equatorial inspirals. Subsequently, we numerically integrate the geodesic equations along this inspiral trajectory and obtain the Boyer--Lindquist coordinates of the small object as a function of time. Finally, we construct a gravitational waveform from the inspiral trajectory using a flat--spacetime wave--emission formula. This scheme, which was developed for the modelling of EMRI systems for LISA, has several useful features: a) the waveforms have been checked against more accurate, Teukolsky--based, waveforms for test-particles on geodesic orbits  and the overlap exceeds \(0.95\) over a large portion of the parameter space \cite{kludge}; b) they are computationally inexpensive; c) conservative self--force corrections to this model have been derived~\cite{cons} for Kerr circular--equatorial orbits at 2PN order. Nonetheless, this model is, at present, incomplete in several ways, e.g., conservative corrections are not yet known for generic orbits; the phase space trajectories are approximate, although they have been matched to Teukolsky based evolutions; and the waveform is constructed from the trajectory using a flat--spacetime wave--emission formula. Nevertheless, this approach should capture the main features of the inspiral waveform accurately. 

The two models differ in their treatment of the plunge and merger phases. The first model, valid for IMBHs of arbitrary spin, employs the transition scheme developed by Ori \& Thorne \cite{amos} to smoothly match the inspiral phase onto the transition to plunge. In the spirit of the ``numerical kludge waveform'', we derive an approximate waveform from this transition trajectory using a flat--spacetime emission formulae applied to the transition orbit and plunge geodesic given by the transition model~\cite{firstpaper}. The inspiral waveform includes the dominant modes \(\ell=m=2\), \(\ell=-m=2\) only and so, for consistency, we included only the same two dominant modes in the expression for the waveform generated during the merger phase.

The second model, used for non--spinning central black holes, uses the EOB prescription~\cite{fai,prcs} to describe the plunge and merger phases. The EOB model is a framework in which the motion of a binary is represented as the effective motion of a test object in a background. Although we use the EOB model only in the non-spinning limit, an extension of the EOB scheme does exist which includes leading--order spin--orbit and spin--spin dynamical effects of a binary system for an ``effective test particle'' moving in a Kerr--type metric \cite{eobs}, and next--to--leading--order spin--orbit couplings \cite{ntlo}. However, it has been recently found \cite{gsve} that it is not straightforward to include higher--order non--spinning PN couplings, such as the 4PN and 5PN adjustable parameters that were recently calibrated to numerical relativity simulations for non-spinning systems~\cite{dna, eobnon}, using these Hamiltonians~\cite{eobs,ntlo}. Additionally, the EOB Hamiltonian in \cite{ntlo} does not reduce to the Hamiltonian of a spinning test particle in the Kerr spacetime. A resolution to this issue was recently proposed in \cite{cha}, in which a canonical Hamiltonian was derived for a spinning test particle in a generic curved spacetime, to linear order in the particle spin. In turn, this canonical Hamiltonian has been used to construct  an improved EOB Hamiltonian \cite{hams}. Additionally, the EOB model has recently been used to model circular--equatorial extreme--mass--ratio inspirals (EMRIs) around spinning supermassive black holes, using fits of various post--Newtonian parameters to Teukolsky--based waveforms~\cite{nic}.  This is the same approach that was previously used with great success for non-spinning black hole systems, e.g., to derive fits for the final mass and spin of a BH after merger that are consistent with NR to about \(\sim 2 \%\) accuracy~\cite{fspin}. We used the non-spinning EOB model only in this work because that model is more mature. However,  comparisons between an IMRI model based on the spinning EOB framework \cite{nic} and the waveforms constructed using the transition model should be pursued in the future.

The final ingredient, again common to both models, is the ringdown waveform. This originates from the distorted Kerr black hole formed after merger, and consists of a superposition of quasinormal modes (QNMs). Each mode has a complex frequency, whose real part is the oscillatory frequency, and whose imaginary part is the inverse of the damping time. Following  Berti, et. al. \cite{rdeq}, and Buonanno, et. al. \cite{rdspin}, we developed a RD model that includes the fundamental mode \((\ell=2,m=2,n=0) \) and two overtones \((n=1,2)\). QNMs always come in pairs which are called ``twin modes''. In general, a mode with a given \((\ell,m)\) will always contain a superposition of two different damped exponentials \cite{rdeq}, one with positive real part of the frequency, and the other with negative real part of the frequency and different damping time.  Omitting these modes would have no serious consequences for non--rotating black holes, but it is conceptually inconsistent for rotating black holes. Furthermore, a single--mode expansion restricts attention to circularly polarized gravitational waves. For full generality, we use a general superposition of modes. The frequencies and damping times of the QNMs depend on the mass and spin after merger of the newly formed Kerr BH and to estimate these we made use of fits for the final mass and spin found in \cite{fai,spif}. The various QNMs are attached to the plunge and merger waveform at the effective light ring by imposing the continuity of the waveform and all necessary higher order time derivatives. 

\begin{figure*}[!htp]
\centerline{
\includegraphics[width=0.5\textwidth, angle=0,clip]{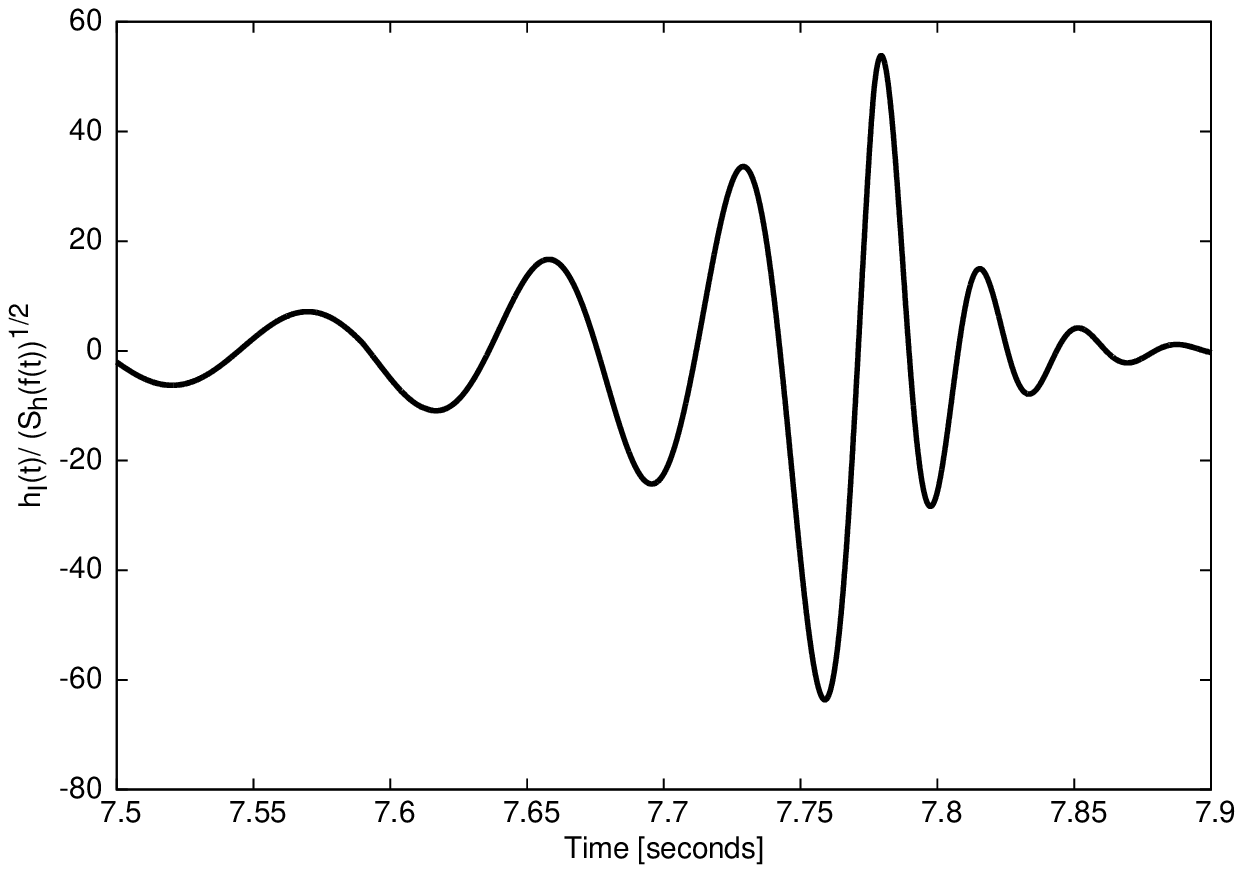}
\includegraphics[width=0.5\textwidth, angle=0,clip]{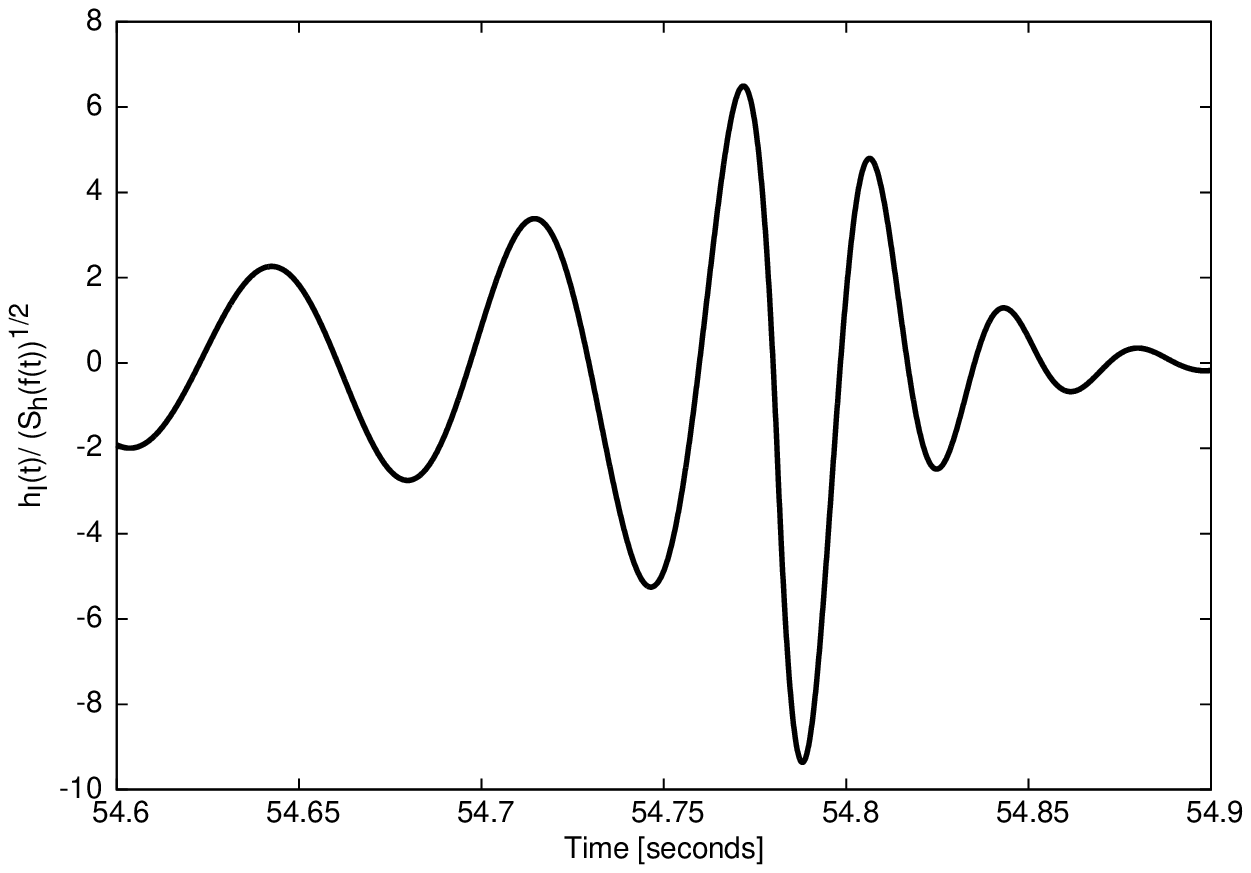}
}
\centerline{
\includegraphics[width=0.5\textwidth, angle=0,clip]{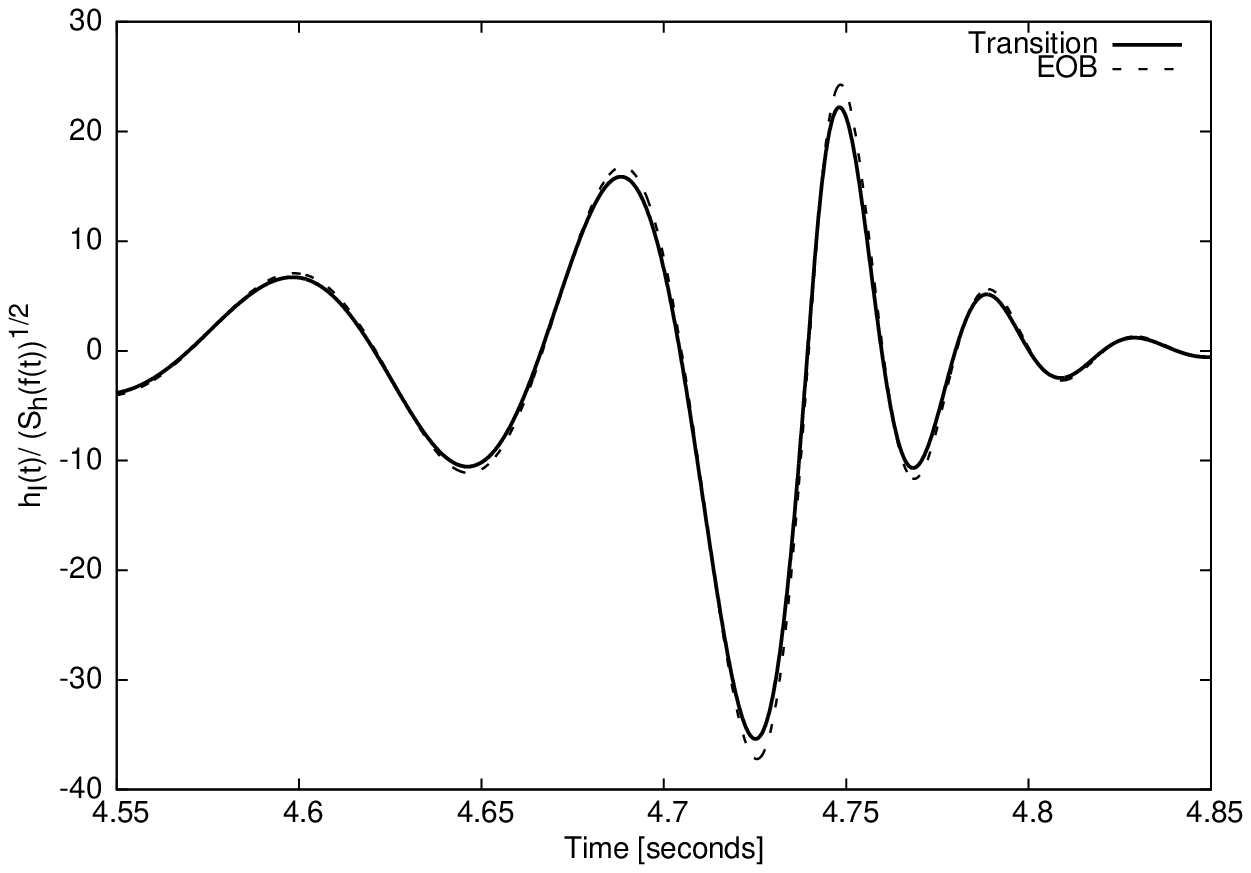}
}
\caption{We show inspiral, merger and ringdown gravitational waveforms for compact objects of mass \(10\,M_{\odot}\) (top left panel and \(1.4 M_{\odot}\) (top right panel),  inspiraling into a \( 500 M_{\odot}\) BH with spin parameter \(q=0.3\). The bottom panel shows the gravitational waveform for a binary system consisting of a compact object of mass \(10\,M_{\odot}\) and a non-spinning ($q=0$) IMBH of mass  \( 500 M_{\odot}\), computed using both the EOB and transition models. In each case, the various extrinsic parameters were chosen randomly. }
\label{figura1}
\end{figure*}

Part of the motivation in developing a waveform model using the transition model of Ori \& Thorne is to enable a more straightforward extension of this present waveform model for circular--equatorial IMRIs  to a waveform model for generic (eccentric--inclined) IMRIs. The ``numerical kludge'' inspiral waveform~ \cite{improved} and the transition model \cite{genor} are both already available for generic orbits. The mass--ratio regime that we explore in these papers has not been as exhaustively explored as the comparable--mass--ratio and extreme--mass--ratio regimes. While the present paper was being prepared, results of a numerical simulation were published which shed light on the gravitational energy and momenta radiated by BH binaries with mass--ratios of 100:1~\cite{carlos}. However, this was the first numerical simulation in the IMRI regime. The schemes described here combine elements of comparable mass and extreme-mass-ratio models in a logical way, but it cannot presently be tested against accurate simulations. As numerical and analytic models improve, these models should be refined. Nonetheless, the models should capture the main features of real IMRI waveforms and therefore be adequate for our purposes, namely to scope out the parameter estimation accuracies that might be achievable with ET. By using two alternative models we can build up confidence in the results. In \cite{firstpaper} we compared the two waveform models for non--spinning systems and demonstrated that the waveforms agreed well in both phase and amplitude, with differences at the level of a few percent. We show sample waveforms computed using the two models in Figure~\ref{figura1}. The close agreement for non-spinning systems is clear from the figure in the lower panel. In this paper, we will now use these models to estimate the precision with which different third-generation detector networks will be able to estimate the parameters of the various binary systems listed above. 

\section{Signal analysis}
\label{s2.2}
In this section we will briefly review the aspects of signal analysis relevant to our calculations. A GW detector is a linear system whose input is the GWs we want to detect and whose output is  a time series, which is a combination of the GW signals and instrumental noise. Any number of coplanar and colocated detectors have a response to GWs that can be written as a linear combination of the responses of two right--angle detectors offset by  \(45^{\circ}\). The output of each of these two equivalent two-arm Michelson detectors can be represented as
\begin{equation}
s_{\alpha}(t) = h_{\alpha}(t) + n_{\alpha}(t), \qquad \alpha = \textrm{I, \,II},
\label{1}
\end{equation}
\noindent where I, II denote the data streams from the two detectors. The detection problem is to distinguish \( h_{\alpha}(t)\) from \( n_{\alpha}(t)\). With the assumptions that (a) each Fourier component  of the noise \({\tilde n_{\alpha}}(f)\) has a Gaussian probability distribution; and (b) the noise is stationary, i.e., the different Fourier components are uncorrelated, the ensemble average of the Fourier components of the noise have the property
\begin{equation}
\label{2}
\langle {\tilde n_\alpha}(f) \, {\tilde n_\beta}(f^\prime)^* \rangle =\frac{1}{2}
\delta(f - f^\prime) S_n(f) \delta_{\alpha \beta}.
\end{equation}
\noindent This relation defines the one-sided spectral density of the instrumental noise, \(S_n(f)\). 

We wish to estimate the parameters of a system from which we have detected gravitational waves. If the expected waveform \(h(t;\theta)\) depends on parameters \(\theta=\{\theta_1,...,\theta_N\}\), we want to construct the posterior probability of the parameters given the data and from that we will find the most probable value of the parameters of the source and their respective uncertainties.

There is a natural inner product on the vector space of signals, which for any two signals \(p_\alpha (t), q_\alpha(t)\), takes the form
\begin{equation}\label{3}
\left( {\bf p} \,|\, {\bf q} \right)
\equiv 2\sum_{\alpha} \int_0^{\infty}\left[ \frac{\tilde p_\alpha^*(f)
\tilde q_\alpha(f) + \tilde p_\alpha(f) \tilde q_\alpha^*(f)}{S_n(f)}\right]
\,\mathrm{d}f.
\end{equation}
\noindent The probability distribution for Gaussian noise, \(n(t)\), is given by
\begin{equation}
\label{4}
p({\bf n}_0) = N \textrm{exp}\left(- \frac{\left( {\bf n}_0\, |\,{\bf n}_0 \right)}{2}\right),
\end{equation}
\noindent where \(N\) is a normalization factor. This relation gives the probability that the actual noise realization is \({\bf n_0}\).  If the source had parameters $\vec{\theta}$, the output would have the form \({\bf s}(t)= {\bf h}(t;\vec{\theta}) + {\bf n}_0(t)\). Using this relation and Equation~(\ref{4}), we obtain the likelihood of observing the output \(s(t)\) if the signal had parameters $\vec{\theta}$
\begin{equation}
\Lambda\left({\bf s}|{\bf \vec{\theta}}\right)= N \textrm{exp}\left(- \frac{\left( {\bf s}_0\,-{\bf h}(\vec{\theta}) |\, {\bf s}_0\,-{\bf h}(\vec{\theta}) \right)}{2}\right).
\label{5}
\end{equation}
\noindent We can use Eq. \eqref{5} in Bayes' theorem, which states that the posterior probability is proportional to the product of the likelihood function and the prior probability. The prior describes the degree of belief, before a measurement is made, that a GW signal in the data would have parameters \(\vec{\theta}\), and the posterior probability describes the corresponding degree of belief after the measurement. Denoting the prior by \(p_0(\vec{\theta})\), the posterior probability for the parameters \(\vec{\theta}\), given the observed output \(s\), is given by \cite{maggiore}
\begin{equation}
\label{6}
p\left( {\bf h}(\vec{\theta}) \,|\, {\bf s} \right) \propto \,p_0(\vec{\theta}) \textrm{exp}\left( \left( {\bf
h}(\vec{\theta})\, |\, {\bf s} \right)- \frac{1}{2}\left( {\bf h}(\vec{\theta})\, |\, {\bf h}(\vec{\theta})
\right)\right).
\end{equation}
\noindent  As \(p\left( {\bf h}(\vec{\theta}) \,|\, {\bf s} \right)\) is a distribution in \(\vec{\theta}\) for a fixed output \({\bf s}\), the term \( \left( {\bf s}\, |\,{\bf s} \right)/2\) can be absorbed into the normalization factor. This probability distribution function describes the information we can extract from the data stream once the prior is specified.

In the absence of an actual data set, we cannot compute this posterior probability distribution, but we can estimate how broad the posterior is likely to be and hence how well we might constrain the parameters. Assuming a flat prior (or a locally-flat prior which would be a reasonable assumption for a signal of high signal-to-noise ratio), the maximum posterior probability is the maximum-likelihood value, which minimizes \(\left( {\bf s-h}\,|\, {\bf s-h} \right) \). This is also the point in parameter space with the highest matched filtering signal--to--noise ratio (SNR)
\begin{equation}
\frac{S}{N}[h(\theta^i)]= \frac{\left( {\bf s}\,|\, {\bf h} \right)}{\sqrt{\left( {\bf h}\,|\, {\bf h} \right)}}.
\label{7}
\end{equation}
\noindent We can expand Eq.~\eqref{6} about this peak, \(\hat{\theta}\) by setting \(\theta^i= \hat{\theta^i} + \Delta \theta^i\) to obtain
\begin{equation}
\label{8}
p(\Delta\theta \, |\,s)\approx\,{\cal N} \, e^{-\frac{1}{2}\Gamma_{ij}\Delta \theta^i
\Delta \theta^j}, \qquad \Gamma_{ij} \equiv \bigg( \frac{\partial {\bf h}}{ \partial \theta^i}\, \bigg| \,
\frac{\partial {\bf h}}{ \partial \theta^j }\bigg)_{|\theta=\hat{\theta}}.
\end{equation}
\noindent \(\Gamma_{ij}\) is the Fisher Information Matrix. This approximation is valid for large SNR. The correlation matrix of the errors \(\Delta \theta^i\) is given by
\begin{equation}
\label{9}
\left< {\Delta \theta^i} {\Delta \theta^j}
 \right>  = (\Gamma^{-1})^{ij}.
\end{equation}
\noindent We shall make use of Eq.~(\ref{9}) to estimate parameter errors for the various binaries described in Section~\ref{s2.1b}. For white noise, i.e., \(S_n(f)= \textrm{const.}\), Parseval's theorem allows us to rewrite the inner product \eqref{3} in the simple form~\cite{cutler}
\begin{equation}
2 \frac{1}{S_n} \sum_{\alpha}\int_{-\infty}^{\infty} \, p_\alpha(t) q_\alpha(t) dt.
\end{equation}
We shall make use of this alternative formulation of the inner product to simplify our analysis.  If we define the ``noise--weighted'' waveform
\begin{equation}\label{10}
\hat h_{\alpha}(t) \equiv   \frac{h_{\alpha}(t)}{\sqrt{S_h\bigl(f(t)\bigr)}}, \qquad f(t) =
\frac{1}{\pi}\frac{\mathrm{d}\phi}{\mathrm{d}t},
\end{equation}
\noindent we can then rewrite the Fisher matrix approximately as \cite{cutler}
\begin{equation}\label{11}
\Gamma_{ab} = 2\sum_{\alpha}\int_0^T{\partial_a \hat h_{\alpha}(t) \partial_b \hat h_{\alpha}(t) \mathrm{d}t} \,, \qquad \mbox{where } \partial_a \equiv \partial/\partial\theta^a.
\end{equation}
\noindent As discussed in \cite{firstpaper}, the binaries we shall consider have relatively low SNR and it is known that in such situations the Fisher Matrix may overestimate measurement accuracies. Vallisneri~\cite{consis} gave a mismatch criterion to determine whether Fisher Matrix results were reliable in a particular context. The idea is to compute the ratio \(r(\theta, {\rm A})\) of the linearized signal amplitude (LSA) likelihood to the exact likelihood. This ratio \(r\) is given by  \cite{consis}
\begin{equation}
\arrowvert \log r(\theta, {\rm A})\arrowvert = \left(\Delta \theta_{j}h_{j} -\Delta h(\theta),\Delta \theta_{k}h_{k} -\Delta h( \theta)\right),
\label{val}
\end{equation} 
\noindent where A represents the signal strength, \( \Delta h(\theta) = h(\theta) - h(\hat \theta)\), with \(\theta= \hat{\theta} + \Delta \theta \), and \(\hat \theta\) the observed location of maximum LSA likelihood for a given experiment. We use the consistency criterion, Eq.~(\ref{val}), to explore the 1--\(\sigma\) likelihood surface predicted by the Fisher Matrix to verify that the mismatch between the LSA and the exact likelihoods is small. Ratios below a fiducial value, say \(\arrowvert \log r(\theta, {\rm A})\arrowvert \sim 0.1\), are considered acceptable~\cite{consis}.

As an example of typical results obtained from this analysis, for \((10M_{\odot}, 100M_{\odot})\) binaries with \(q=0.3\), we found that systems near the lower quartile of the distribution had \(\arrowvert \log r(\theta, {\rm A})\arrowvert \sim 0.2\). From this threshold onwards, the ratio \(r\) decreases gradually so that at the upper quartile of the distribution we obtain \(\arrowvert \log r(\theta, {\rm A})\arrowvert \sim 0.04\). This indicates that the results we get from the Fisher Matrix should be a reasonable estimate of the measurement precisions achievable for IMRIs using ET. Our results may be somewhat optimistic and at some point should be verified using Monte Carlo simulations to recover the full posterior probability distributions. Such an exercise is beyond the scope of this present work. Nonetheless, the results should be fairly accurate, and we will see in Section~\ref{s2.3} that, where comparisons can be made, our results are in good accord with previous results that have been derived for other types of binary and using different waveform models.  

The Fisher matrix for a network of detectors is given by the sum of the individual Fisher Matrices for each detector. When modelling the response of the various detectors, it is necessary to account for their relative positions on the surface of the Earth, as the corresponding time delays are what allow source triangulation. In computing the waveforms used in the Fisher Matrix, we found it convenient to use two different timesteps in order to separately resolve the (slow) inspiral and (fast) merger/ringdown phases. We checked that varying the choice of timesteps and the location of the transition between the two timesteps did not significantly affect the results for any of the systems considered. We will discuss the convergence of the Fisher Matrices further in Section~\ref{s2.3}.

\section{Parameter estimation error results}
\label{s2.3}
The parameter space of the signals we shall consider is 10 dimensional. Four of these are intrinsic parameters, namely \(\ln m, \ln M, q, t_0\). The other six are extrinsic parameters. We summarize the physical meaning of all the parameters in Table \ref{tableparams}.

To explore parameter estimation errors using the inverse  Fisher Matrix we fix the values of the intrinsic parameters of the source, and carry out a Monte Carlo simulation over possible values for the extrinsic parameters. The parameter errors scale with the SNR of the source as 1/SNR, so we can quote results at a fixed SNR and these may be easily extrapolated to other SNRs. We chose a reference SNR of 30, as this would represent a very robust detection for which the Fisher Matrix prediction is likely to be a good estimate. To obtain this reference result, we first compute the Fisher Matrix for a source at a fixed distance, $D=6.6348$Gpc, and obtain the SNR at that distance from the expression
\begin{equation}\label{12}
{\rm SNR}^2= 2\sum_{\alpha=I,II}\int_{t_{\rm init}}^{t_{\rm LSO}}
\hat h_{\alpha}^2(t)dt.
\end{equation}  
\noindent We then multiply the errors estimated from the Fisher Matrix by (SNR\(/30\)) to normalise to the reference SNR of 30. We assume that the observation starts when the GWs from the inspiral reach a frequency of 5Hz and finishes when the RD waveform is no longer contributing to the SNR. 

\begin{table}[thb]
\centerline{$\begin{array}{c|l}\hline\hline
 \ln m      & \text {mass of compact object (CO)}   \\
 \ln M         & \text {mass of IMBH}\\
 q         & \text{magnitude of (specific) spin
    angular momentum of IMBH} \\
t_0           & \text{time at which orbital frequency sweeps through a reference value} \\
\phi_0 &  \text{initial phase of CO orbit}      \\
\theta_S  &  \text{source sky colatitude in an ecliptic--based system }  \\
\phi_S   & \text{source sky azimuth in an ecliptic--based system}  \\
 \theta_K   & \text{direction  of IMBH spin (colatitude)}  \\
\phi_K       &  \text{direction of IMBH spin (azimuth)}  \\
\ln D          & \text{distance to  source}\\
\hline\hline
\end{array}$}
\caption{\protect\footnotesize
This table shows the physical meaning of the parameters used in our model. The angles ($\theta_S$,\,$\phi_S$) and ($\theta_K$,\,$\phi_K$) are defined in a fixed ecliptic--based coordinate system.}
\label{tableparams}
\end{table}

\subsection{Dependence of parameter estimation errors on system parameters}
We consider the twelve different binary systems described in Section~\ref{s2.1b}. For the sources with $q=0.3$, $0.9$ we present results computed from the transition model for the waveform. For the non-spinning systems, $q=0$, we present results from both the transition model and the EOB model. The Tables~\ref{M100m10}--\ref{M500m14} list the mean, standard  deviation, median and lower and upper quartiles of the distribution of the Fisher Matrix errors computed in the Monte Carlo simulation. There is one table for each of the four mass combinations we consider and all results are computed at the fixed reference SNR of 30. These results assume that the detection is made using the optimistic 3 ET network configuration, and hence should be considered as upper bounds on the accuracy with which we may be able to measure the various parameters.

\begin{figure*}[!htp]
\centerline{
\includegraphics[width=0.55\textwidth, angle=0,clip]{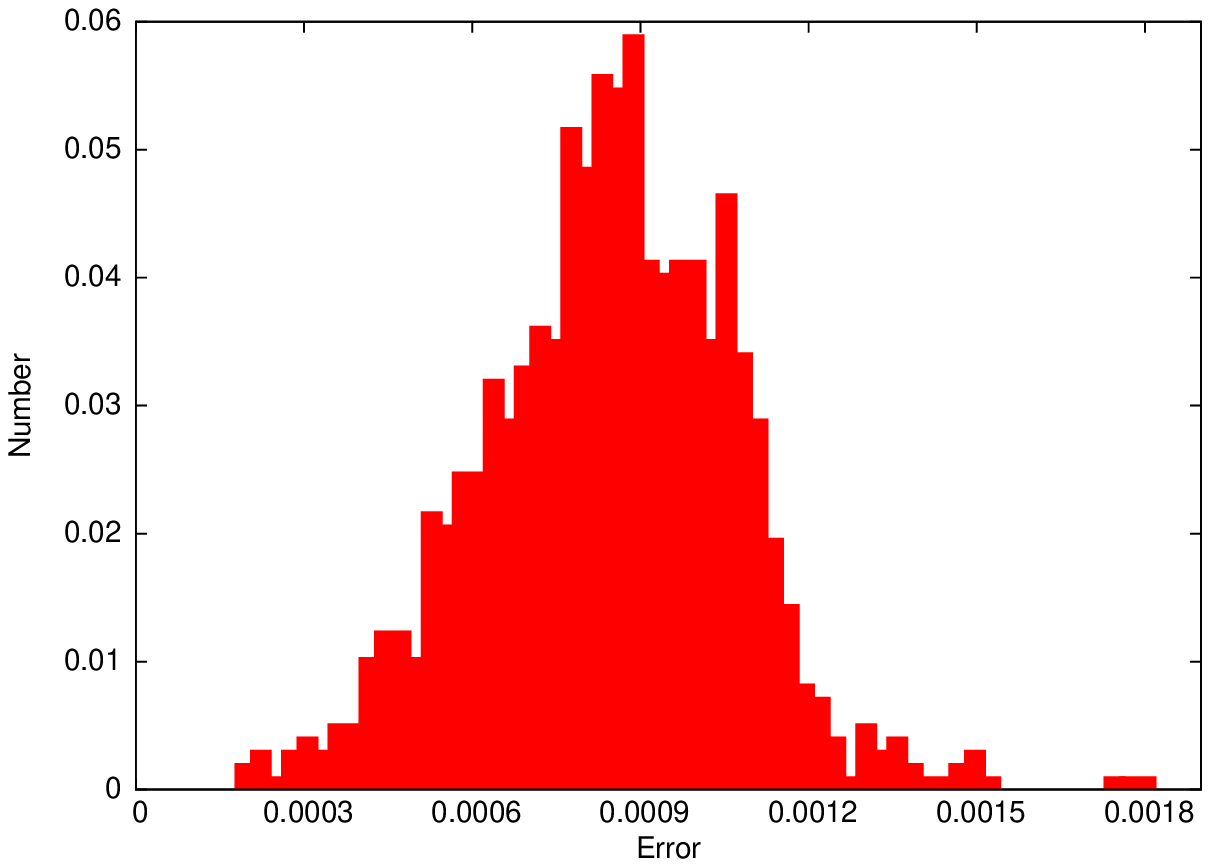}
\includegraphics[width=0.55\textwidth, angle=0,clip]{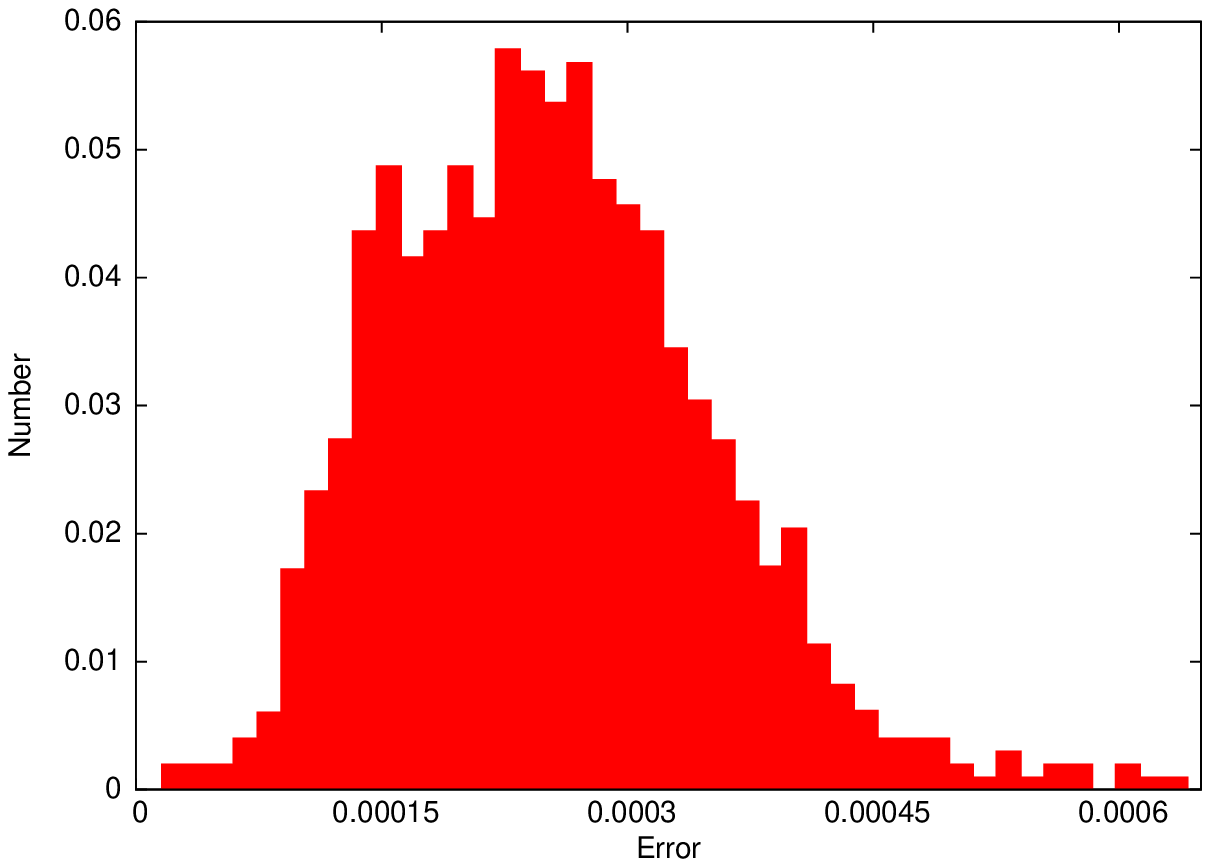}
}
\centerline{
\includegraphics[width=0.55\textwidth, angle=0,clip]{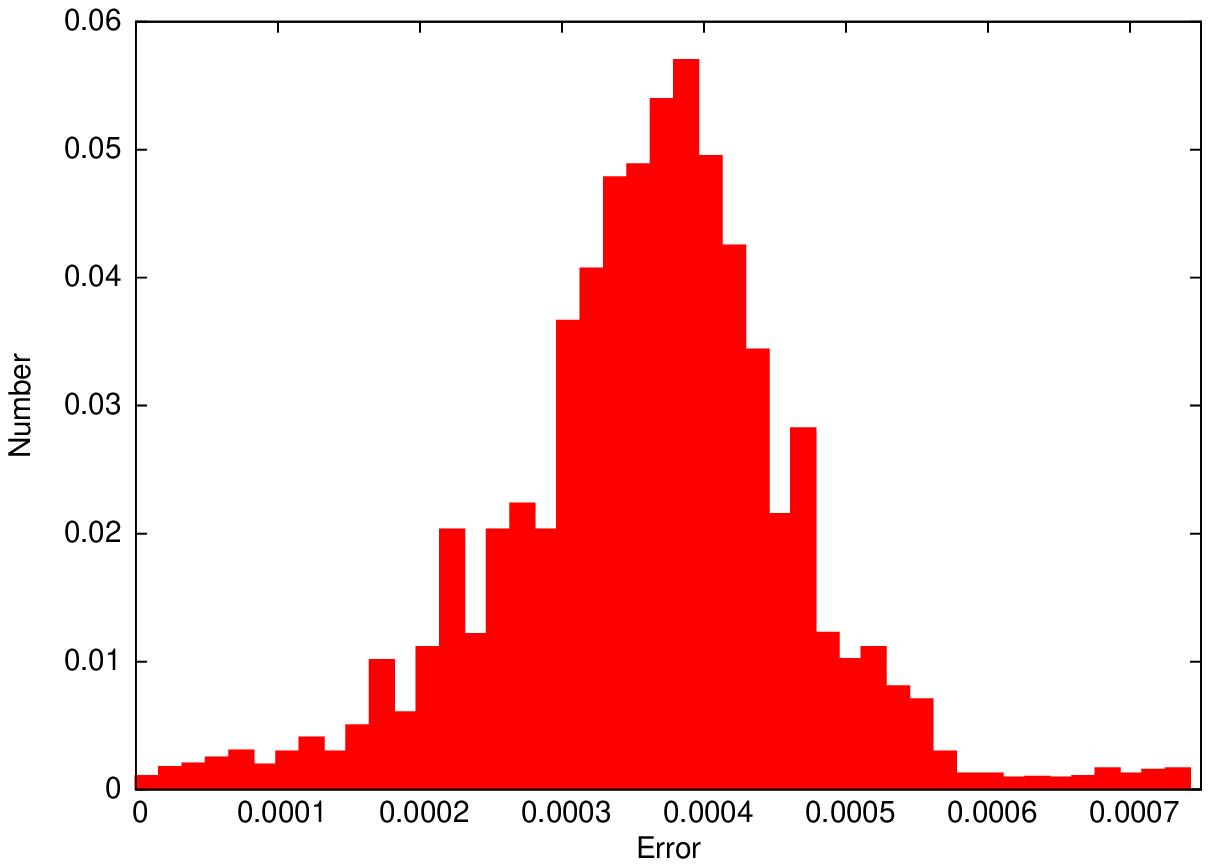}
\includegraphics[width=0.55\textwidth, angle=0,clip]{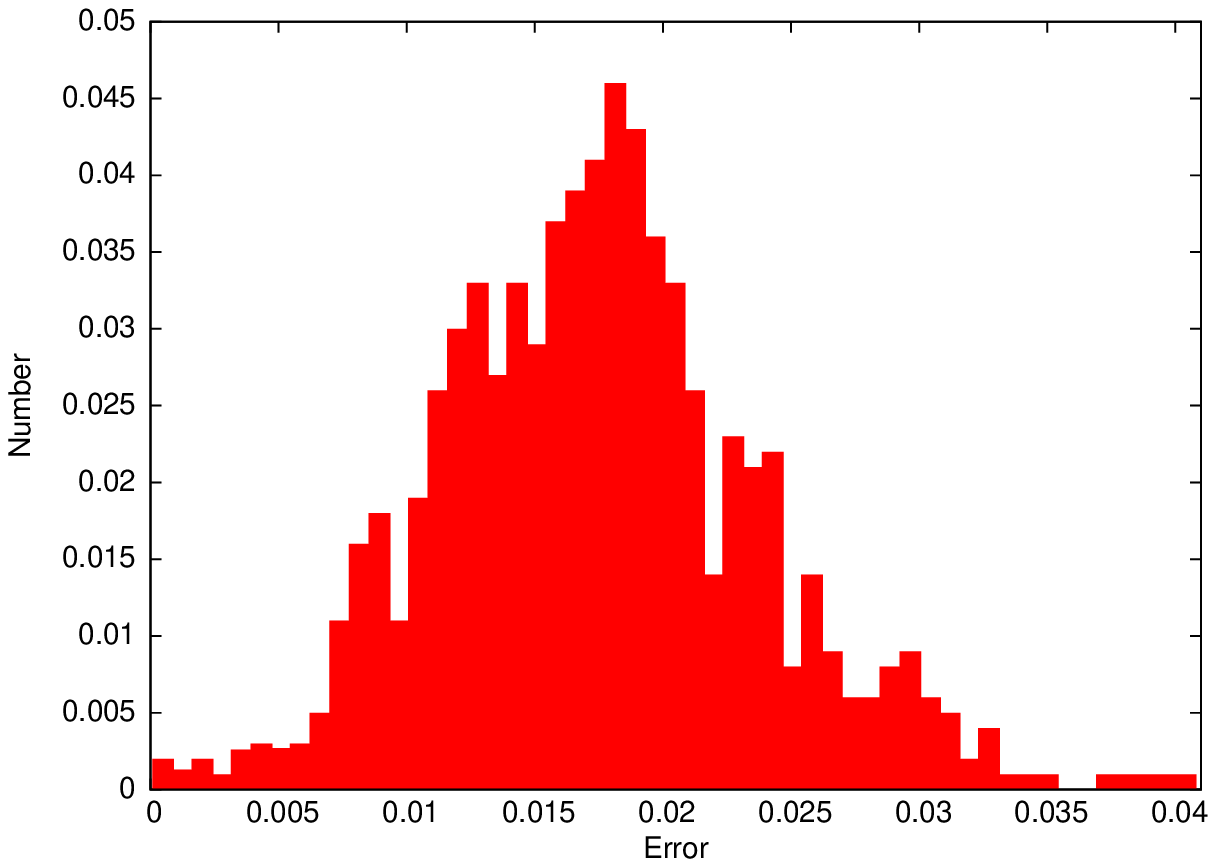}
}
\centerline{
\includegraphics[width=0.55\textwidth, angle=0,clip]{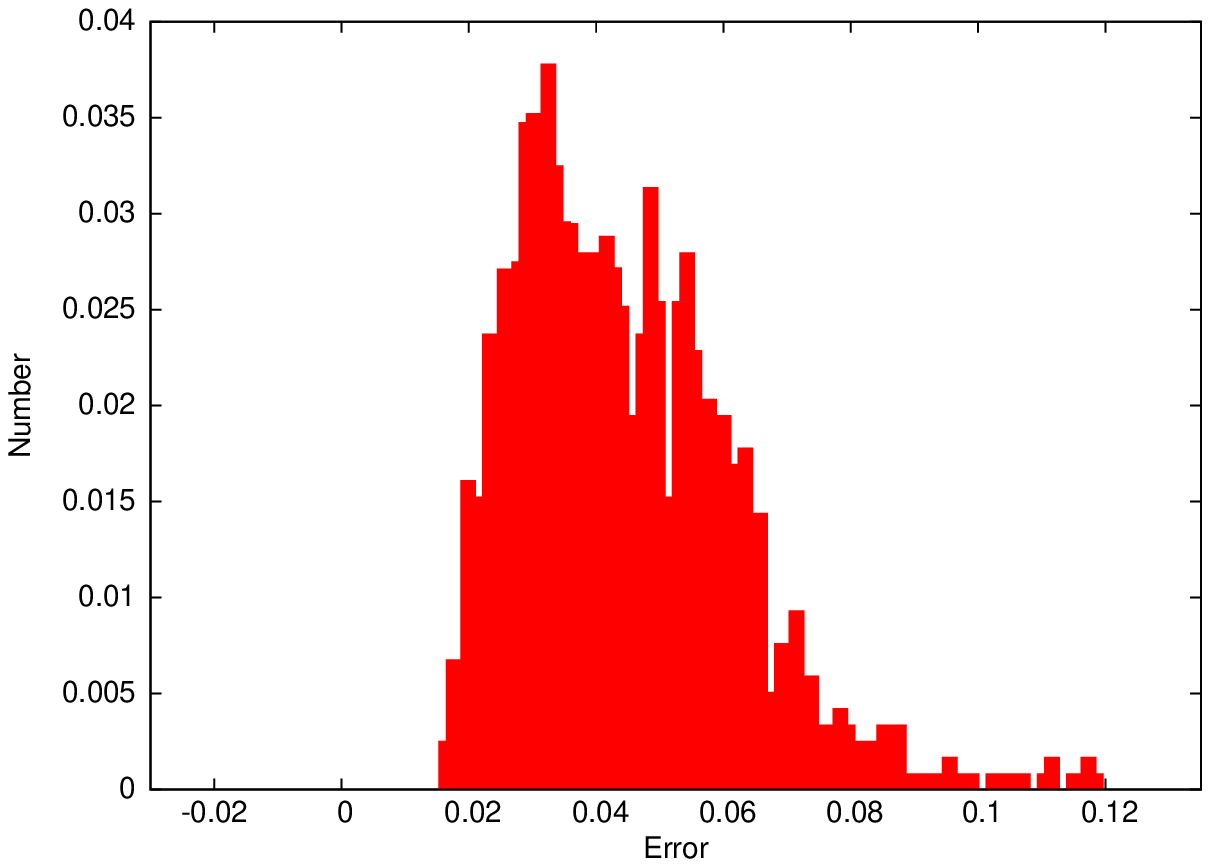}
\includegraphics[width=0.55\textwidth, angle=0,clip]{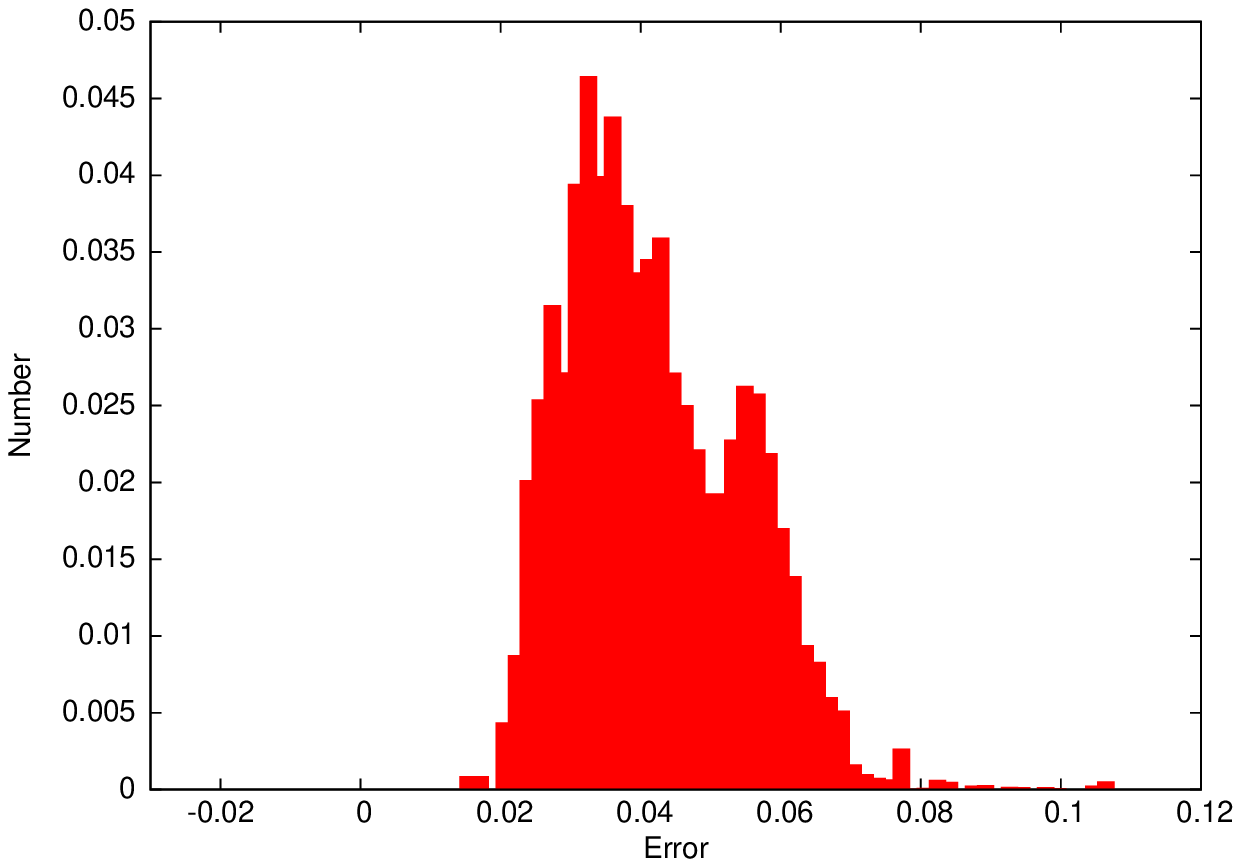}
}
\caption{Distribution of parameter measurement error estimates computed in the Monte Carlo simulations. We show results for the system with  \(m=10M_{\odot}\),  \(M=100M_{\odot}\), \(q=0.9\). The panels show the error distributions in the following order, top row, from left to right, \(\Delta(\ln m)\), \( \Delta(\ln M) \); middle row, \( \Delta q \), \(\Delta t_0  \); bottom row, \( \Delta \theta_S \), \( \Delta \phi_S \).}
\label{hist}
\end{figure*}

In Figure~\ref{hist} we show histograms of the error estimates for several parameters found in the Monte Carlo simulation for the \(m=10M_{\odot}\), \(M=100M_{\odot}\), \(q=0.9\) system.  This figure, and Table~\ref{M100m10}, indicate that a network of 3 ETs might be able to determine the location of the source in the sky to an accuracy of \(\sim10^{-3}\) steradians, i.e., \(\sim 4\) square degrees. This estimate is the statistical mean of the Monte Carlo distribution at SNR of 30 assuming the optimistic detector network of three ETs in the geographic locations of Virgo, LIGO Livingston and Perth, Australia (AIGO). 

Recent work has shown that for the existing LIGO-Virgo detector network, assuming a uniform distribution of sources across the sky, at a network SNR of around 15, \(50\%\) of inspiral sources should be located within 23 sq--degs (best case) at the \(95\%\) confidence level. For burst sources, without any knowledge of the waveform, \(50\%\) of the sources could be localized within 50 sq--degs (worst case) at an SNR of 10, but this can be reduced to 8 sq--degs if predicted waveforms are available \cite{ares}. These estimates are similar for the initial or advanced detectors \cite{ares}, at a fixed SNR. The inclusion of an additional detector in the Southern Hemisphere, such as AIGO, can further improve these values as it contributes a longer baseline, additional energy flux, extended signal space, and breaks the plane--degeneracy formed by the three detector network of LIGO Livingston, LIGO Hanford and Virgo~\cite{ares}. 

Another recent independent study \cite{fairhurst}  presented results for the accuracy with which sources could be localized with a network of GW detectors, using only timing information in the various detectors. Fairhurst \cite{fairhurst} found that increasing the number of detectors at different sites increases both the absolute number of observable sources, and greatly increases the fraction of sources that can be well localized. For instance, at a fixed SNR of 8, and using the advanced detector network comprising the two LIGO detectors at Hanford (HH) and LIGO Livingston (L), no sources can be localized within 20 sq--degs. Adding another detector at a different site, e.g., a LIGO detector in Australia (A), or Advanced Virgo (V), or the Japanese LCGT detector (J), the networks HLA, HHLV, HHJL, can localize up to \(50\%\) of the signals within 20 sq--degs, and the loudest signals within 5 sq--degs.  In all these results, it was found that for the networks involving an Australian detector, the peak of the localization distributions occurs between 5 and 10 sq--degs. Our results imply that we can determine the location of the source in the sky to an accuracy of \(\sim 12\) sq--degs when we renormalise to a network SNR of 8. This estimate is the statistical median value obtained from the Monte Carlo simulations. Our results are therefore in good accord with existing estimates of the angular resolution that could be achieved by the advanced detector network.  Furthermore, Table~\ref{M100m10} shows that for non--spinning systems, we should resolve the plunge time \(t_0\) to within \(\sim 30\) ms when normalised to an SNR of 10. This estimate is also in good accord with the results obtained independently by Ajith et al. \cite{tnought} and Luna et al. \cite{luna}.

The Monte Carlo simulations also show that we expect to determine the total mass of  \(m=10M_{\odot}\,, M=100M_{\odot}\) binaries to an accuracy of \(\sim 0.1\%\) for non--spinning systems when we renomalise the results to an SNR of 8. This estimate is in good accord with previous estimates for related systems. In~\cite{fwwd} it was estimated that using the same 3ET detector network that we have considered in this paper, it should be possible to determine the total mass of a non=spinning \(m=23M_{\odot}\,, M=100M_{\odot}\) binary (mass ratio \(\eta = 0.16\)) to an accuracy of \(\sim 0.1\%\) at a network SNR of 8. This estimate was obtained using the phenomenological waveform model described in \cite{ajith}, which includes the inspiral, merger and ringdown for non-spinning, comparable-mass binaries  in a consistent way.

\begin{figure*}[!htp]
\begin{tabular}{cc}
\includegraphics[width=0.5\textwidth, angle=0,clip]{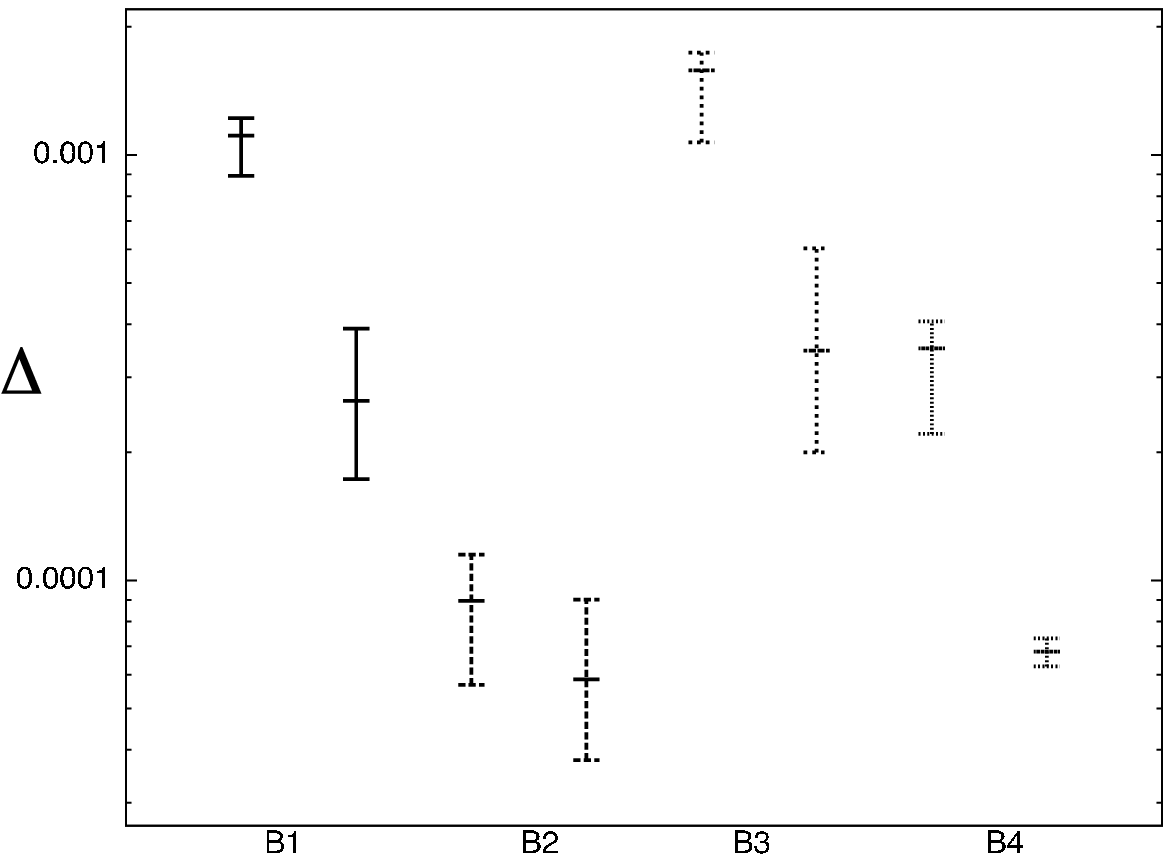}&
\includegraphics[width=0.5\textwidth, angle=0,clip]{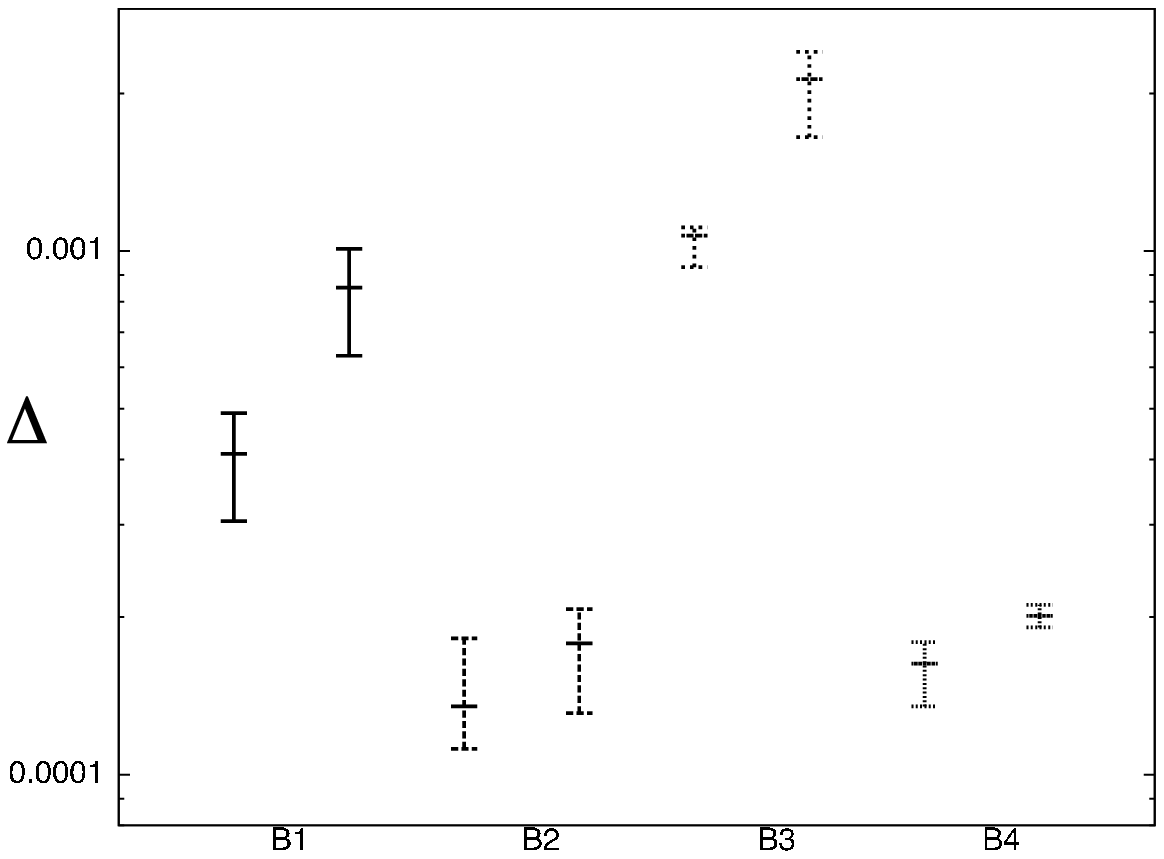}
\end{tabular}
\caption{Left panel: expected fractional measurement errors, \(\Delta\), in the mass of the CO and the mass of the IMBH for a fixed value of the IMBH spin parameter \(q=0.3\). There is a pair of candlesticks for each of the four  binary systems. For each binary, the candlestick representing the CO's mass error estimate is to the left of that for the IMBH's mass error estimate. 
The error bars indicate the lower and upper quartiles found in the Monte Carlo simulation.  Right panel: expected measurement errors in the IMBH's spin parameter for two different values for the spin parameter \(q\). For each system, the candlestick corresponding to a central IMBH with \(q=0.3\) is to the right of that associated with the error estimate for a system with \(q=0.9\).  The continuous, long--dashed, dashed and dotted lines represent systems B1: \(m=10M_{\odot} \,, M=100M_{\odot}\), B2: \(m=1.4M_{\odot} \,, M=100M_{\odot}\), B3: \(m=10M_{\odot} \,, M=500M_{\odot}\), and B4: \(m=1.4M_{\odot} \,,  M=500M_{\odot}\), respectively.}
\label{whiskm}
\end{figure*}

Figure~\ref{whiskm} shows the accuracy with which we expect to determine the masses of the small CO and the IMBH for the spinning systems, as well as the IMBH's spin parameter \(q\). This figure shows that for binaries that stay the longest in band, namely the \(m=1.4M_{\odot} \,,  M=100M_{\odot}\), \(q=0.3\) system ( \(\sim 291 \) seconds in band), we can determine the masses of the CO and the IMBH to an accuracy of \(0.01\%\), \(0.005\%\), respectively, at the reference SNR of 30. The accuracy gradually decreases for the shorter--lived events. For instance, for binaries with \(m=10M_{\odot} \,,  M=500M_{\odot}\), \(q=0.3\), which stay in band about \(\sim 8.3 \) seconds, we expect to measure the CO's and IMBH's masses to an accuracy of only \(\sim0.15\%\), \(\sim0.05\%\) respectively, at the same SNR of 30. 
 
We expect to determine the spin parameter more accurately for rapidly rotating binaries. For instance, the estimated accuracy with which we can measure the IMBH's spin for binaries with \(m=10M_{\odot} \,,  M=100M_{\odot}\), \(q=0.9\) is \(\sim 0.05 \% \). This is about a factor of 2 better than for binaries with the same component masses but with spin \(q=0.3\).  We would expect this trend since, for more rapidly spinning systems, the small CO comes much closer to the outer horizon of the Kerr IMBH before merging, and this is the regime where the CO can more strongly feel the effects of the IMBH's spin. We note that the {\it percentage} error in the spin is quite similar in the two cases. This is consistent with this understanding --- the spin is measured more precisely when it is larger because it has a correspondingly greater effect on the system. For less massive COs, the inspiral proceeds more slowly and the early inspiral has a correspondingly greater importance in the estimation of the system parameters. For such systems, we might expect the precision of spin determination to be les sensitive to the system spin, and this is borne out by our results --- the spin precision is comparable for $q=0.9$ and $q=0.3$ in the $1.4M_{\odot} +100M_{\odot}$ and $1.4M_{\odot} +500M_{\odot}$ systems.

In general, we expect that the precision with which we can measure the parameters of a binary depends on the number of GW cycles that are observed. Therefore, we might expect the precision to depend on the source masses as, from best to worst,  \(1.4M_{\odot}\, + 100M_{\odot}\),  \(1.4M_{\odot}\, + 500M_{\odot}\), \(10M_{\odot}\, + 100M_{\odot}\), \(10M_{\odot}\, + 500M_{\odot}\).  Tables~\ref{M100m10}--\ref{M500m14}, and Figure~\ref{whiskm} confirms this expectation.  These Tables also show that the precision improves for more rapidly spinning IMBHs. This is because, for greater spins, the inspiral phase evolution lasts longer and so there are more cycles of information in the waveform. The SNR of the spinning systems will also be larger at a given distance, as compared with slowly spinning IMBHs, so these systems can be seen further away. This does not affect the results quoted here, which are normalised to fixed SNR=30.  Tables~\ref{M100m10}--\ref{M500m14} also confirm that our two independent waveform models make predictions for the parameter estimation errors for non-spinning systems that are consistent to better than ten per cent.

Finally, we note that the consistency of the distributions of Figure~\ref{hist}, in the sense that they are smooth with few outliers, is an indication that the results are convergent. It is well known that Fisher Matrices encountered in parameter estimation calculations for GW sources can have very large condition numbers. However, the results we present here were obtained from Fisher Matrices that exhibited convergence over at least two orders of magnitude in the offsets used to compute the numerical waveform derivatives. The inverse matrices were computed using an LU decomposition, and we verified that the inverse Fisher matrices were also convergent to \(\lesssim 10\%\) over an order of magnitude in the numerical offsets. We also found that the offsets required for the various network configurations were consistent, as we would expect, since the convergence of the FM should depend on the intrinsic waveform, rather than the choice of network.

\subsection{Dependence of parameter estimation errors on network configuration}
\begin{figure*}[!htp]
\centerline{
\includegraphics[width=0.5\textwidth, angle=0,clip]{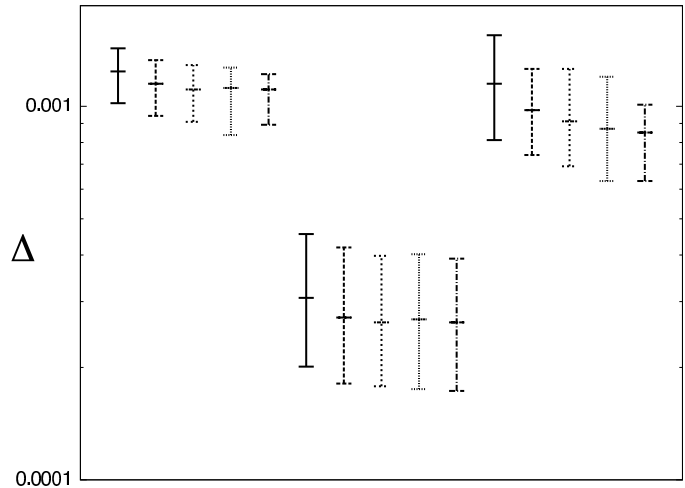}
}
\caption{Arranged from left to right in the panel, we show the expected measurement errors in the CO's mass, IMBH mass and IMBH spin parameter, as a function of the network configuration for binary systems with masses \(10M_{\odot}\, + 100M_{\odot}\) and IMBH spin \(q=0.3\). In each group the candlesticks are arranged in order C1--C5 from left to right. As before, the CO and IMBH mass errors are quoted as fractional errors, i.e., these are the errors in $\ln(m)$, $\ln(M)$.}
\label{whiskc}
\end{figure*}

We shall now consider the more modest network configurations C1--C4 described earlier, which consist of combinations of ET's and right--angle detectors at 2 or 3 sites chosen from Virgo, LIGO Livingston and Perth (Australia). These results are shown in Tables~\ref{conf1}--\ref{conf4}\footnote[1]{Note that only Table~\ref{conf1} is included in the main text. Tables~\ref{conf2}--\ref{conf4} may be found in Appendix A.} for the four different mass combinations, but with the central IMBH spin fixed at \(q=0.3\). As expected, we find that a single ET is sufficient for accurate intrinsic parameter determination. This is because the signature of the intrinsic parameters is encoded in the phase evolution which can be accurately measured by a single instrument. After normalising the parameter errors to a fixed network SNR of 30, Figure~\ref{whiskc} shows clearly that the accuracy with which we can measure the intrinsic parameters of a binary is not dramatically changed by changes in the network configuration. 
 
However, it is not possible to constrain the location of the source in the sky nor the luminosity distance using a single detector. This is also expected, as the determination of the extrinsic parameters comes primarily from the time--delays between the arrival of the signal in different detectors. It appears that  the inclusion of an additional right--angle detector (configuration C2) is enough to constrain the source sky position and luminosity distance to moderate precision \(\sim 30\) sq--degs, \(\sim 15\%\) for a  \( 10M_{\odot}\) BH IMRI into a \( 100M_{\odot}\) IMBH with spin \(q=0.3\) at SNR of 30. Replacing this right--angle detector by another ET at the same location (configuration C3) enhances the determination of these parameters to \(\sim 12\) sq--degs, \(\sim10\%\), respectively. If we consider a configuration consisting of 1 ET, plus two right--angle detectors (configuration C4) these estimates are further improved to \(\sim 11\) sq--degs, \(\sim 10 \%\), respectively. These two estimates are rather close to the precision we expect to obtain using a 3 ET network, namely, \(\sim 8\) sq--degs, \( \sim 10\%\), respectively. 

Moreover, these results also show that the distributions become narrower for the more complex configurations, i.e., there is less variation in the precisions with which we can determine the extrinsic parameters as we randomise over the source position and orientation. This is also expected, as a more complex network should provide more complete sky coverage. We can compare our results for the luminosity distance  in the non--spinning limit with those quoted in Table I of \cite{fwwd}, which were given at a network SNR of 8. Renormalising our results for the non--spinning systems to an SNR of 8, we estimate that network C2 would be able to measure the luminosity distance to an accuracy of \(\sim40\%\). For configuration C3, we would be able to determine the luminosity distance to an accuracy of \(30\%\) and for C4, this improves marginally to \(\sim28\%\). This estimate is roughly the same for the 3ET detector network (C5). These estimates are in good accord with Table I of \cite{fwwd}, which were computed for a \( \sim 20M_{\odot}\, + 100M_{\odot}\) binary system using a comparable-mass inspiral-merger-ringdown waveform model. This comparison provides further support that even if the IMRI models used in this paper are not accurate enough to be used for detection of IMRI systems, they do still give reasonable results and can be used for parameter estimation studies to illustrate the potential scientific impact of the ET.

In summary, a single ET will be enough to measure the intrinsic parameters of a system but will not be sufficient to accurately reconstruct the extrinsic parameters. Extrinsic parameter determination is possible with the addition of one right-angle detector, but a 2 ET configuration will improve the accuracy with which we can measure the source's sky position and luminosity distance by roughly a  factor of \(\sim 2\).  A more optimistic configuration consisting of a single ET and two right-angle detectors upgraded to ET's sensitivity will generate results that are competitive with the highly optimistic three ET network. This configuration (C3) might be more realistic, since it will have lower overall costs, although it is not simply an upgrade of existing sites as we are assuming $10$km scale right-angle interferometers.

Tables~\ref{M100m10}--\ref{conf4} show that GW observations will produce valuable astrophysical results. A network of 2 ETs operating in coincidence will be able to measure accurately the masses of two merging black holes, and less accurately the sky position and luminosity distance at which the merger is taking place. The results in paper I~\cite{firstpaper} suggested that a a few tens to a few hundreds of IMRI events could be detected per year by an ET network, provided that IMBHs are relatively abundant in globular clusters, and the various mechanisms leading to CO capture --- secular Kozai resonance, binary exchange processes, gravitational radiation and three and four--body interactions --- are efficient at starting IMRIs. The ET network would be able to achieve good parameter determinations for all the systems that are detected, which will have important scientific implications.

One important question will be whether the IMBHs detected have been formed in clusters, or are primordial, population III, black holes formed in the early Universe. The precision achievable with network C4, of  \(\sim 11\) sq--degs in sky position and \(\sim10\%\) in luminosity distance will not be enough to determine the host galaxy of an IMRI uniquely, let alone if it is occurring in a cluster within that galaxy. If the the ET low-frequency cut-off does extend down to $1$Hz, these precisions improve to \(\sim 1\) sq--degs, and \(\sim 7\%\), but even that is unlikely to be enough. However, primordial IMBHs are likely to be hosted in dwarf galaxies today, and the stellar density is sufficiently low that is very unlikely that IMRIs could occur there~\cite{etgair}. Thus, it would be a reasonable assumption that any IMRIs detected would be from cluster IMBHs and not population III black holes. The measured masses and IMBH spins thus provide constraints on the cluster IMBH population, and their evolution.

IMRI systems could be detected up to redshifts as high as  \(z \sim 6\), if the system has redshifted masses \([10 + 100]M_{\odot}\) and IMBH spin \(q=0.9\), assuming a 3ET network with a frequency cut--off of 1Hz \cite{firstpaper}. At such redshifts, the existence of light black holes, no matter how they formed, provides some constraints on the hierarchical assembly of structure. If IMBHs exist at $z \sim 6$, they will inevitably grow and may be progenitors of some fraction of today's supermassive black holes. IMRI detections with ET could therefore be important for cosmology, which relies on the ability to confidently say that an event is at high redshift, which we have demonstrated is possible.

A redshift $z\sim 6$ is probably not high enough to provide strong constraints on structure growth, but ET might also detect IMBHs when they merge with comparable mass IMBHs, and these can be seen out to redshifts $z\gtrsim 10$~\cite{fwwd,etgair}. Again, the IMBHs involved in such mergers could form either in clusters or in the early Universe. As many as a few tens of mergers involving primordial IMBHs could be detected by ET~\cite{fwwd} in the light-seed scenario. The number of comparable mass mergers of IMBHs formed in clusters could be much higher, as many as several thousand~\cite{etgair}, but this is very dependent on the efficiency of IMBH formation in the cluster environment. As mentioned above, the IMRIs detected by ET are most likely to be occurring in clusters and thus provide a direct constraint on the existence, number density and properties of IMBHs in clusters. This information can be used to estimate how many of the observed IMBH-IMBH mergers might be coming from the cluster channel and hence what fraction of events might be primordial. IMRI observations will therefore also have indirect applications to our understanding of structure formation.

The Monte Carlo results we present in this paper are the first results for IMRI sources detectable by ET to appear in the literature. The results must be taken with some caution, because of the approximations in the waveform models that have been made, as discussed earlier. However, these results should be a reasonable guide to the likely order-of-magnitude of the parameter measurement errors that will be achievable. The precise astrophysical implications will depend primarily on the number of events that are seen, which is very difficult to predict given current uncertainties in the astrophysics of IMBHs, and indeed their very existence.

\begin{table}[thb]
\begin{tabular}{|c|c|c|c|c|c|c|c|c|c|c|c|}
\hline\multicolumn{2}{|c|}{}&\multicolumn{10}{c|}{Statistics of distribution for error \(\Delta X\) in parameter \(X=\)}\\\cline{3-12}
\multicolumn{2}{|c|}{Model}&$\ln(m)$&$\ln(M)$&$q$&$t_0$&$\phi_0$&$\theta_S$&$\phi_S$&$\theta_K$&$\phi_K$&$\ln(D)$\\\hline
&Mean           &8.31e-4        &2.81e-4        &3.97e-4        &1.66e-2        &1.37e-1        &4.07e-2        &5.09e-2        &8.19e-2        &1.01e-1        &1.01e-1\\\cline{2-12}
&St. Dev.       &2.27e-4        &1.47e-4        &1.63e-4        &9.17e-3        &4.98e-2        &1.71e-2        &2.74e-2        &6.59e-2        &6.92e-2        &6.64e-2\\\cline{2-12}
q=0.9&L.Qt.     &6.86e-4        &1.74e-4        &3.05e-4        &8.69e-3        &1.24e-1        &2.92e-2        &3.46e-2        &4.04e-2        &4.95e-2        &5.15e-2\\\cline{2-12}
&Med.           &8.45e-4        &2.60e-4        &4.10e-4        &1.50e-2        &1.68e-1        &4.16e-2        &4.62e-2        &6.54e-2        &7.09e-2        &8.01e-2\\\cline{2-12}
&U. Qt.         &9.59e-4        &3.44e-4        &4.90e-4        &2.23e-2        &1.96e-1        &5.49e-2        &6.06e-2        &9.55e-2        &1.10e-1        &1.28e-1\\\hline

&Mean           &1.09e-3        &2.96e-4        &8.76e-4        &1.51e-2        &1.13e-1        &4.28e-2        &5.42e-2        &7.80e-2        &9.77e-2        &1.05e-1\\\cline{2-12}
&St. Dev.       &3.68e-4        &1.22e-4        &3.13e-4        &8.39e-3        &3.55e-2        &1.54e-2        &2.48e-2        &5.17e-2        &4.35e-2        &7.14e-2\\\cline{2-12}
q=0.3&L.Qt.     &8.93e-4        &1.73e-4        &6.31e-4        &1.12e-2        &8.95e-2        &2.95e-2        &3.54e-2        &3.93e-2        &4.64e-2        &5.53e-2\\\cline{2-12}
&Med.           &1.11e-3        &2.64e-4        &8.51e-4        &1.66e-2        &1.29e-1        &4.16e-2        &5.04e-2        &5.73e-2        &7.45e-2        &8.64e-2\\\cline{2-12}
&U. Qt.         &1.22e-3        &3.91e-4        &1.01e-3        &2.34e-2        &1.59e-1        &5.22e-2        &6.96e-2        &8.64e-2        &1.24e-1        &1.39e-1\\\hline

&Mean           &4.46e-4 &2.29e-4 &N/A &1.41e-2 &1.21e-1 &4.32e-2 &5.95e-2 &8.53e-2 &1.07e-1 &1.19e-1\\\cline{2-12}
&St. Dev.       &2.04e-4 &1.39e-4 &N/A &5.62e-3 &5.17e-2 &1.61e-2 &3.23e-2 &6.51e-2 &7.94e-2 &7.16e-2\\\cline{2-12}
q=0&L.Qt.       &3.10e-4 &1.70e-4 &N/A &9.33e-3 &8.49e-2 &3.09e-2 &3.54e-2 &4.00e-2 &4.85e-2 &5.97e-2\\\cline{2-12}
&Med.           &4.91e-4 &2.51e-4 &N/A &1.32e-2 &1.06e-1 &4.16e-2 &5.07e-2 &6.48e-2 &7.43e-2 &8.76e-2\\\cline{2-12}
&U. Qt.         &6.18e-4 &3.89e-4 &N/A &1.69e-2 &1.48e-1 &5.32e-2 &7.42e-2 &8.79e-2 &1.19e-1 &1.45e-1\\\hline

&Mean           &3.95e-4  &2.11e-4 &N/A &1.15e-2 &1.23e-1 &4.50e-2 &5.79e-2 &8.48e-2 &9.77e-2 &1.16e-1\\\cline{2-12}
&St. Dev.       &1.94e-4  &1.29e-4 &N/A &6.02e-3 &4.87e-2 &1.93e-2 &2.98e-2 &5.89e-2 &6.26e-2 &6.37e-2\\\cline{2-12}
EOB&L.Qt.       &2.90e-4  &1.55e-4 &N/A &7.24e-3 &9.17e-2 &3.08e-2 &3.46e-2 &4.19e-2 &4.74e-2 &5.94e-2\\\cline{2-12}
&Med.           &4.01e-4  &2.25e-4 &N/A &1.07e-2 &1.17e-1 &4.13e-2 &4.91e-2 &6.56e-2 &7.76e-2 &8.45e-2\\\cline{2-12}
&U. Qt.         &5.32e-4  &2.99e-4 &N/A &1.29e-2 &1.48e-1 &5.49e-2 &7.21e-2 &8.73e-2 &1.01e-1 &1.50e-1\\\hline

\end{tabular}
\caption{Summary of Monte Carlo results for parameter estimation errors.  We show the mean, standard deviation, median and quartiles of the distribution  of the error in each parameter.   Results are given for a \(m=10M_{\odot}\)  CO inspiralling into a \(M=100M_{\odot}\) IMBH for various choices of the IMBH spin, \(q\), computed using the transition model waveform. We also show results for $q=0$ computed using the EOB waveform model.}
\label{M100m10}
\end{table}

\begin{table}[thb]
\begin{tabular}{|c|c|c|c|c|c|c|c|c|c|c|c|}
\hline\multicolumn{2}{|c|}{}&\multicolumn{10}{c|}{Statistics of distribution  for error \(\Delta X\) in parameter \(X=\)}\\\cline{3-12}
\multicolumn{2}{|c|}{Model}&$\ln(m)$&$\ln(M)$&$q$&$t_0$&$\phi_0$&$\theta_S$&$\phi_S$&$\theta_K$&$\phi_K$&$\ln(D)$\\\hline
&Mean      &9.53e-6     &4.30e-6        &1.32e-4        &1.94e-2        &1.44e-1        &4.18e-2        &4.67e-2        &8.32e-2        &8.90e-2        &9.39e-2\\\cline{2-12}
&St. Dev.  &4.35e-6     &2.17e-6        &4.10e-5        &7.26e-3        &6.21e-2        &1.41e-2        &1.57e-2        &6.51e-2        &6.62e-2        &6.02e-2\\\cline{2-12}
q=0.9&L.Qt.&6.47e-6     &3.12e-6        &1.12e-4        &1.31e-2        &9.18e-2        &2.95e-2        &3.16e-2        &4.27e-2        &5.28e-2        &5.08e-2\\\cline{2-12}
&Med.      &9.22e-6     &4.51e-6        &1.35e-4        &1.86e-2        &1.51e-1        &4.06e-2        &4.39e-2        &6.79e-2        &7.26e-2        &6.90e-2\\\cline{2-12}
&U. Qt.    &1.17e-5     &5.94e-6        &1.82e-4        &2.51e-2        &2.09e-1        &5.24e-2        &5.94e-2        &9.44e-2        &9.98e-2        &1.12e-1\\\hline

&Mean      &9.41e-5     &5.99e-5        &1.80e-4        &1.65e-2        &1.36e-1        &3.97e-2        &4.63e-2        &7.91e-2        &9.15e-2        &9.96e-2\\\cline{2-12}
&St. Dev.  &4.24e-5     &3.13e-5        &6.86e-5        &8.11e-3        &6.83e-2        &1.25e-2        &1.84e-2        &5.48e-2        &6.27e-2        &5.84e-2\\\cline{2-12}
q=0.3&L.Qt.&5.68e-5     &3.78e-5        &1.31e-4        &1.28e-2        &1.05e-1        &3.05e-2        &3.18e-2        &3.86e-2        &4.91e-2        &5.11e-2\\\cline{2-12}
&Med.      &8.96e-5     &5.85e-5        &1.78e-4        &1.69e-2        &1.37e-1        &3.96e-2        &4.59e-2        &5.74e-2        &7.81e-2        &7.24e-2\\\cline{2-12}
&U. Qt.    &1.15e-4     &9.01e-5        &2.07e-4        &2.08e-2        &1.67e-1        &5.17e-2        &5.98e-2        &1.04e-1        &9.95e-2        &1.20e-1\\\hline

&Mean    &8.43e-5       &6.03e-5        &N/A            &1.34e-2        &1.15e-1        &4.19e-2        &5.16e-2        &8.72e-2        &9.41e-2        &1.00e-1\\\cline{2-12}
&St. Dev.&3.98e-5       &3.02e-5        &N/A            &7.52e-3        &4.55e-2        &1.51e-2        &2.50e-2        &6.50e-2        &7.49e-2        &6.40e-2\\\cline{2-12}
q=0&L.Qt.&7.21e-5       &3.82e-5        &N/A            &9.33e-3        &8.58e-2        &2.91e-2        &3.38e-2        &4.08e-2        &4.41e-2        &5.67e-2\\\cline{2-12}
&Med.    &8.71e-5       &5.85e-5        &N/A            &1.28e-2        &1.09e-1        &4.07e-2        &4.57e-2        &6.09e-2        &7.38e-2        &6.92e-2\\\cline{2-12}
&U. Qt.  &1.05e-4       &7.75e-5        &N/A            &1.77e-2        &1.67e-1        &5.28e-2        &6.47e-2        &9.46e-2        &1.09e-1        &1.11e-1\\\hline

&Mean    &8.61e-5       &5.65e-5        &N/A            &1.20e-2        &1.24e-1        &4.30e-2        &5.64e-2        &8.93e-2        &9.65e-2        &1.04e-1\\\cline{2-12}
&St. Dev.&3.16e-5       &2.15e-5        &N/A            &5.72e-3        &4.34e-2        &1.86e-2        &2.71e-2        &6.97e-2        &7.48e-2        &6.87e-2\\\cline{2-12}
EOB&L.Qt.&7.08e-5       &3.69e-5        &N/A            &1.09e-2        &8.95e-2        &3.01e-2        &3.28e-2        &4.13e-2        &4.98e-2        &5.36e-2\\\cline{2-12}
&Med.    &8.51e-5       &5.21e-5        &N/A            &1.25e-2        &9.59e-2        &4.16e-2        &4.68e-2        &5.93e-2        &7.66e-2        &7.56e-2\\\cline{2-12}
&U. Qt.  &1.02e-4       &7.18e-5        &N/A            &1.38e-2        &1.71e-1        &5.39e-2        &7.14e-2        &9.72e-2        &1.13e-1        &1.31e-1\\\hline
\end{tabular}
\caption{As Table~\ref{M100m10}, but now for binary systems with a CO of mass \(m=1.4M_{\odot}\).}
\label{M100m14}
\end{table}

\begin{table}[thb]
\begin{tabular}{|c|c|c|c|c|c|c|c|c|c|c|c|}
\hline\multicolumn{2}{|c|}{}&\multicolumn{10}{c|}{Statistics of distribution for error \(\Delta X\) in parameter \(X=\)}\\\cline{3-12}
\multicolumn{2}{|c|}{Model}&$\ln(m)$&$\ln(M)$&$q$&$t_0$&$\phi_0$&$\theta_S$&$\phi_S$&$\theta_K$&$\phi_K$&$\ln(D)$\\\hline
&Mean      &1.33e-3     &5.10e-4        &1.01e-3        &1.41e-2        &1.21e-1        &4.47e-2        &5.59e-2        &9.86e-2        &1.07e-1        &1.15e-1\\\cline{2-12}
&St. Dev.  &6.04e-4     &1.66e-4        &2.79e-4        &4.77e-3        &4.87e-2        &1.64e-2        &2.42e-2        &6.80e-2        &7.47e-2        &4.66e-2\\\cline{2-12}
q=0.9&L.Qt.&9.53e-4     &4.39e-4        &9.32e-4        &1.15e-2        &8.21e-2        &3.02e-2        &3.41e-2        &4.86e-2        &5.68e-2        &6.80e-2\\\cline{2-12}
&Med.      &1.26e-3     &5.08e-4        &1.07e-3        &1.57e-2        &1.11e-1        &4.16e-2        &4.89e-2        &6.86e-2        &6.83e-2        &8.91e-2\\\cline{2-12}
&U. Qt.    &1.67e-3     &5.93e-4        &1.11e-3        &1.84e-2        &1.58e-1        &5.62e-2        &7.04e-2        &1.14e-1        &1.27e-1        &1.39e-1\\\hline
&Mean       &1.41e-3     &3.63e-4       &2.09e-3        &1.19e-2        &1.05e-1        &3.95e-2        &5.12e-2        &9.57e-2        &1.08e-1        &1.12e-1\\\cline{2-12}
&St. Dev.   &6.39e-4     &1.87e-4       &5.72e-4        &5.07e-3        &5.20e-2        &1.19e-2        &2.03e-2        &6.74e-2        &8.75e-2        &7.67e-2\\\cline{2-12}
q=0.3&L.Qt. &1.07e-3     &2.00e-4       &1.65e-3        &7.41e-3        &7.83e-2        &2.96e-2        &3.43e-2        &5.74e-2        &5.75e-2        &5.93e-2\\\cline{2-12}
&Med.       &1.58e-3     &3.47e-4       &2.13e-3        &1.24e-2        &9.38e-2        &3.86e-2        &4.79e-2        &6.91e-2        &8.79e-2        &8.82e-2\\\cline{2-12}
&U. Qt.     &1.74e-3     &6.03e-4       &2.40e-3        &1.41e-2        &1.22e-1        &4.92e-2        &6.51e-2        &1.14e-1        &1.24e-1        &1.56e-1\\\hline
&Mean      &4.06e-3     &1.54e-3        &N/A            &1.01e-2        &9.86e-2        &4.36e-2        &5.54e-2        &1.00e-1        &1.14e-1        &1.25e-1\\\cline{2-12}
&St. Dev.  &2.47e-3     &6.58e-4        &N/A            &5.69e-3        &5.05e-2        &1.63e-2        &2.65e-2        &6.31e-2        &7.58e-2        &8.10e-2\\\cline{2-12}
q=0&L.Qt.  &2.95e-3     &1.20e-3        &N/A            &8.14e-3        &7.71e-2        &2.96e-2        &3.53e-2        &4.50e-2        &5.25e-2        &5.82e-2\\\cline{2-12}
&Med.      &4.09e-3     &1.51e-3        &N/A            &1.17e-2        &1.04e-1        &4.06e-2        &4.99e-2        &6.84e-2        &8.31e-2        &9.11e-2\\\cline{2-12}
&U. Qt.    &6.21e-3     &1.99e-3        &N/A            &1.35e-2        &1.24e-1        &5.55e-2        &7.01e-2        &1.13e-1        &1.21e-1        &1.71e-1\\\hline
&Mean     &3.59e-3      &1.53e-3        &N/A            &1.02e-2        &1.00e-1        &4.63e-2        &5.72e-2        &9.69e-2        &1.07e-1        &1.31e-1\\\cline{2-12}
&St. Dev. &1.07e-3      &5.89e-4        &N/A            &5.46e-3        &3.46e-2        &2.01e-2        &2.91e-2        &7.56e-2        &8.94e-2        &8.29e-2\\\cline{2-12}
EOB&L. Qt.&2.88e-3      &1.04e-3        &N/A            &7.94e-3        &8.77e-2        &3.02e-2        &3.44e-2        &4.22e-2        &5.62e-2        &6.31e-2\\\cline{2-12}
&Med.     &3.62e-3      &1.39e-3        &N/A            &1.10e-2        &9.92e-2        &4.16e-2        &5.24e-2        &6.56e-2        &8.12e-2        &9.92e-2\\\cline{2-12}
&U. Qt.   &4.09e-3      &1.75e-3        &N/A            &1.38e-2        &1.26e-1        &5.71e-2        &7.42e-2        &1.01e-1        &1.19e-1        &1.79e-1\\\hline
\end{tabular}
\caption{ As Table~\ref{M100m10}, but now for binary systems with a central IMBH of mass \(M=500M_{\odot}\). }
\label{M500m10}
\end{table}

\begin{table}[thb]
\begin{tabular}{|c|c|c|c|c|c|c|c|c|c|c|c|}
\hline\multicolumn{2}{|c|}{}&\multicolumn{10}{c|}{Statistics of distribution for error \(\Delta X\) in parameter \(X=\)}\\\cline{3-12}
\multicolumn{2}{|c|}{Model}&$\ln(m)$&$\ln(M)$&$q$&$t_0$&$\phi_0$&$\theta_S$&$\phi_S$&$\theta_K$&$\phi_K$&$\ln(D)$\\\hline
&Mean      &2.69e-4     &8.71e-5        &1.38e-4        &1.41e-2        &1.14e-1        &4.34e-2        &5.77e-2        &9.61e-2        &1.18e-1        &9.44e-2\\\cline{2-12}
&St. Dev.  &1.14e-4     &3.21e-5        &5.90e-5        &4.57e-3        &4.41e-2        &1.68e-2        &2.85e-2        &7.54e-2        &7.70e-2        &5.45e-2\\\cline{2-12}
q=0.9&L.Qt.&1.95e-4     &7.41e-5        &1.35e-4        &1.23e-2        &9.79e-2        &2.92e-2        &3.56e-2        &4.79e-2        &5.08e-2        &4.94e-2\\\cline{2-12} 
&Med.      &2.78e-4     &8.91e-5        &1.63e-4        &1.47e-2        &1.21e-1        &4.26e-2        &5.02e-2        &6.51e-2        &8.18e-2        &6.81e-2\\\cline{2-12}
&U. Qt.    &3.49e-4     &1.01e-4        &1.79e-4        &1.73e-2        &1.44e-1        &5.49e-2        &7.18e-2        &1.16e-1        &1.50e-1        &1.21e-1\\\hline
&Mean      &3.17e-4     &6.82e-5        &2.03e-4        &1.14e-2        &9.02e-2        &4.02e-2        &5.75e-2        &8.59e-2        &1.10e-1        &9.68e-2\\\cline{2-12}
&St. Dev.  &1.03e-4     &8.51e-6        &4.31e-5        &5.22e-3        &5.32e-2        &1.28e-2        &2.84e-2        &6.22e-2        &9.12e-2        &6.10e-2\\\cline{2-12}
q=0.3&L.Qt.&2.21e-4     &6.28e-5        &1.91e-4        &1.02e-2        &7.02e-2        &3.02e-2        &3.23e-2        &4.48e-2        &5.25e-2        &4.91e-2\\\cline{2-12}
&Med.      &3.51e-4     &6.80e-5        &2.01e-4        &1.17e-2        &9.34e-2        &4.16e-2        &5.17e-2        &6.14e-2        &8.22e-2        &7.39e-2\\\cline{2-12}
&U. Qt.    &4.06e-4     &7.31e-5        &2.11e-4        &1.65e-2        &1.24e-1        &5.09e-2        &6.97e-2        &1.06e-1        &1.50e-1        &1.21e-1\\\hline
&Mean    &2.71e-4       &1.07e-4        &N/A            &1.03e-2        &9.18e-2        &4.33e-2        &5.69e-2        &8.35e-2        &1.10e-1        &1.11e-1\\\cline{2-12}
&St. Dev.&5.44e-5       &2.10e-5        &N/A            &3.09e-3        &5.43e-2        &1.58e-2        &2.81e-2        &6.35e-2        &8.71e-2        &7.23e-2\\\cline{2-12}
q=0&L.Qt.&2.51e-4       &1.00e-4        &N/A            &6.31e-3        &6.32e-2        &3.09e-2        &3.46e-2        &4.17e-2        &4.75e-2        &5.09e-2\\\cline{2-12}
&Med.    &2.63e-4       &1.17e-4        &N/A            &1.09e-2        &9.41e-2        &4.27e-2        &5.12e-2        &5.84e-2        &7.81e-2        &8.70e-2\\\cline{2-12}
&U. Qt.  &2.95e-4       &1.23e-4        &N/A            &1.25e-2        &1.17e-1        &5.49e-2        &7.21e-2        &1.03e-1        &1.50e-1        &1.30e-1\\\hline
&Mean      &2.87e-4     &1.25e-4        &N/A            &1.07e-2        &9.46e-2        &4.75e-2        &5.70e-2        &8.53e-2        &1.07e-1        &1.23e-1\\\cline{2-12}
&St. Dev.  &8.23e-5     &3.33e-5        &N/A            &4.37e-3        &3.22e-2        &1.66e-2        &2.94e-2        &6.72e-2        &7.61e-2        &8.49e-2\\\cline{2-12}
EOB&L.Qt.  &2.29e-4     &1.02e-4        &N/A            &7.94e-3        &7.06e-2        &3.02e-2        &3.54e-2        &4.36e-2        &4.79e-2        &6.02e-2\\\cline{2-12}  
&Med.      &2.75e-4     &1.20e-4        &N/A            &1.04e-2        &9.18e-2        &4.16e-2        &4.82e-2        &5.91e-2        &7.53e-2        &8.75e-2\\\cline{2-12}
&U. Qt.    &3.09e-4     &1.35e-4        &N/A            &1.23e-2        &1.16e-1        &5.85e-2        &7.41e-2        &9.98e-2        &1.36e-1        &1.34e-1\\\hline
\end{tabular}
\caption{As Table~\ref{M100m14}, but now for binary systems with a central IMBH of mass \(M=500M_{\odot}\). }
\label{M500m14}
\end{table}

\begin{table}[thb]
\begin{tabular}{|c|c|c|c|c|c|c|c|c|c|c|c|}
\hline\multicolumn{2}{|c|}{}&\multicolumn{10}{c|}{Statistics of distribution for error \(\Delta X\) in parameter \(X=\)}\\\cline{3-12}
\multicolumn{2}{|c|}{Model}&$\ln(m)$&$\ln(M)$&$q$&$t_0$&$\phi_0$&$\theta_S$&$\phi_S$&$\theta_K$&$\phi_K$&$\ln(D)$\\\hline
&Mean           &1.23e-3        &3.45e-4        &1.15e-3        &2.04e-2        &1.62e-1        &2.2995        &2.7421          &3.7537        &4.0738         &4.2386\\\cline{2-12}
&St. Dev.       &4.19e-4        &1.99e-4        &4.23e-4        &9.71e-3        &5.90e-2        &3.4104        &4.0524          &3.9352        &4.0112         &4.1351\\\cline{2-12}
C1&L.Qt.        &1.02e-3        &2.01e-4        &8.12e-4        &1.41e-2        &1.23e-1        &0.0866        &0.1018          &1.0189        &1.1825         &1.2491\\\cline{2-12}
&Med.           &1.24e-3        &3.07e-4        &1.15e-3        &1.99e-2        &1.62e-1        &0.6820        &0.8174          &2.2722        &2.6118         &2.6707\\\cline{2-12}
&U. Qt.         &1.43e-3        &4.55e-4        &1.55e-3        &2.81e-2        &1.84e-1        &2.9151        &3.3087          &4.9711        &5.3026         &5.7026\\\hline

&Mean           &1.11e-3        &3.13e-4        &9.12e-4        &1.69e-2        &1.34e-1        &6.19e-2        &7.61e-2        &1.50e-1        &1.88e-1        &1.76e-1\\\cline{2-12}
&St. Dev.       &4.84e-4        &1.84e-4        &3.57e-4        &8.43e-3        &4.35e-2        &3.77e-2        &4.62e-2        &1.28e-1        &1.76e-1        &1.54e-1\\\cline{2-12}
C2&L.Qt.        &9.43e-4        &1.81e-4        &7.41e-4        &1.23e-2        &1.08e-1        &3.31e-2        &3.76e-2        &6.60e-2        &6.34e-2        &9.05e-2\\\cline{2-12}
&Med.           &1.15e-3        &2.72e-4        &9.77e-4        &1.82e-2        &1.45e-1        &5.18e-2        &6.22e-2        &1.01e-1        &1.28e-1        &1.47e-1\\\cline{2-12}
&U. Qt.         &1.33e-3        &4.19e-4        &1.26e-3        &2.63e-2        &1.75e-1        &7.79e-2        &1.07e-1        &2.03e-1        &2.48e-1        &2.74e-1\\\hline

&Mean           &1.07e-3        &3.04e-4        &8.87e-4        &1.66e-2        &1.29e-1        &4.56e-2        &5.81e-2        &8.47e-2        &1.14e-1        &1.11e-1\\\cline{2-12}
&St. Dev.       &3.73e-4        &1.75e-4        &3.59e-4        &8.01e-3        &4.12e-2        &2.15e-2        &3.27e-2        &6.78e-2        &9.85e-2        &8.35e-2\\\cline{2-12}
C3&L.Qt.        &9.09e-4        &1.78e-4        &6.91e-4        &1.12e-2        &1.08e-1        &2.89e-2        &3.24e-2        &4.07e-2        &4.92e-2        &5.81e-2\\\cline{2-12}
&Med.           &1.11e-3        &2.64e-4        &9.12e-4        &1.77e-2        &1.45e-1        &4.29e-2        &4.92e-2        &6.07e-2        &7.83e-2        &8.73e-2\\\cline{2-12}
&U. Qt.         &1.29e-3        &3.98e-4        &1.26e-3        &2.63e-2        &1.75e-1        &5.95e-2        &7.20e-2        &9.78e-2        &1.49e-1        &1.45e-1\\\hline

&Mean           &1.05e-3        &3.01e-4        &8.79e-4        &1.58e-2        &1.18e-1        &4.47e-2        &5.61e-2        &8.24e-2        &1.14e-1        &1.12e-1\\\cline{2-12}
&St. Dev.       &4.27e-4        &1.78e-4        &3.56e-4        &8.14e-3        &4.38e-2        &2.09e-2        &2.87e-2        &6.38e-2        &9.81e-2        &8.07e-2\\\cline{2-12}
C4&L.Qt.        &8.38e-4        &1.75e-4        &6.31e-4        &1.12e-2        &1.04e-1        &3.21e-2        &3.54e-2        &4.46e-2        &4.82e-2        &5.38e-2\\\cline{2-12}
&Med.           &1.12e-3        &2.69e-4        &8.71e-4        &1.77e-2        &1.45e-1        &4.06e-2        &5.02e-2        &6.91e-2        &7.71e-2        &8.52e-2\\\cline{2-12}
&U. Qt.         &1.27e-3        &4.02e-4        &1.20e-3        &2.39e-2        &1.69e-1        &5.45e-2        &6.91e-2        &9.54e-2        &1.33e-1        &1.40e-1\\\hline

&Mean           &1.09e-3        &2.96e-4        &8.76e-4        &1.51e-2        &1.13e-1        &4.28e-2        &5.42e-2        &7.80e-2        &9.77e-2        &1.05e-1\\\cline{2-12}
&St. Dev.       &3.68e-4        &1.22e-4        &3.13e-4        &8.39e-3        &3.55e-2        &1.54e-2        &2.48e-2        &5.17e-2        &4.35e-2        &7.14e-2\\\cline{2-12}
C5&L.Qt.        &8.93e-4        &1.73e-4        &6.31e-4        &1.12e-2        &8.39e-2        &2.95e-2        &3.54e-2        &3.93e-2        &4.64e-2        &5.53e-2\\\cline{2-12}
&Med.           &1.11e-3        &2.64e-4        &8.51e-4        &1.66e-2        &1.29e-1        &4.16e-2        &5.04e-2        &5.73e-2        &7.45e-2        &8.64e-2\\\cline{2-12}
&U. Qt.         &1.22e-3        &3.91e-4        &1.01e-3        &2.34e-2        &1.59e-1        &5.22e-2        &6.96e-2        &8.64e-2        &1.24e-1        &1.39e-1\\\hline
\end{tabular}
\caption{As Table~\ref{M100m10}, but for binary systems with a central IMBH of spin parameter \(q=0.3\) and assuming four alternative configurations for the detector network, C1-C4, as described in Section~\ref{s2.1}. Configuration C5 is the network of three ETs which has been used for all results elsewhere in this paper.  }
\label{conf1}
\end{table}

\section{Conclusions}
\label{s2.4}
In this paper, we have used the gravitational waveform models for IMRIs developed in paper I of this series~\cite{firstpaper} to estimate the precision with which the Einstein Telescope will be able to determine the parameters of circular--equatorial IMRIs. We have presented results for a set of twelve ``typical'' systems, comprising four different combinations of component masses --- $1.4M_{\odot} + 100M_{\odot}$,  $1.4M_{\odot} + 500M_{\odot}$, $10M_{\odot}+100M_{\odot}$, and $10M_{\odot}+500M_{\odot}$ --- and three different IMBH spins --- \(q=0,0.3,0.9\). For the non--spinning systems, we have compared the results from the transition model and the EOB waveform model and found that these models make predictions that are consistent to better than ten percent. This final check provides confidence in these results.  

We have also explored how the accuracy of parameter determination depends on the configuration of the detector network using the ``ET--B'' noise curve and assuming a cut--off frequency of 5Hz. We have shown that a single ET is sufficient to accurately determine the intrinsic parameters of these systems. However,  a network of detectors is required to obtain accurate estimates of the extrinsic parameters. A network of 2 ETs should be sufficient to measure the source's sky position and luminosity distance to accuracies of \(\sim 12\) sq--degs and \(\sim10\%\), respectively, for a BH IMRI into a \( 100M_{\odot}\) IMBH with spin \(q=0.3\). A more sophisticated network comprising 1 ET and two right-angle detectors would have comparable precisions. The results for the latter networks are comparable to the precisions that could be achieved with a 3 ET network, which are \(\sim 8\) sq--degs, \(10\%\), respectively (at a source SNR of 30). Any one of these ET networks would simultaneously be able to constrain the BH and IMBH masses and the IMBH spin magnitude to fractional accuracies of \(\sim 10^{-3},\,  10^{-3.5}\, \textrm{and } 10^{-3}\), respectively. The amount of variation in the parameter precision over random choices of the source location and orientation also decreases for more complex network configurations. A 3 ET network is a highly optimistic assumption about a future third--generation GW detector network, but our results indicate that a more modest network comprising one ET and right-angle interferometers in LIGO Livingston and Perth can recover parameters to almost the same precision. This network would have lower associated costs and might therefore be more feasible. Our results should be regarded as conservative in the sense that using a lower low-frequency cut-off, or assuming a different ET design, e.g., the xylophone configuration~\cite{Hild:2009}, could improve the various results quoted in these papers. At present we have made use of the ``ET--B'' noise curve \cite{etb}, but a study of the potential applications of ET using both lower cut--off frequencies and more optimistic designs should be carried out in the future.

If IMRIs are detected, there will be several important scientific payoffs. The detection of an IMBH as a GW source would provide the first robust proof of the existence of black holes in the intermediate mass range. As discussed in the previous section, IMRIs are most likely to occur in globular clusters, thus IMRI observations will provide information about whether IMBHs form at all in these stellar environments, and whether they remain in the clusters rather than being ejected. The masses, spins and abundance of IMBHs detected as IMRIs will provide constraints on the formation efficiency and evolution of globular cluster systems. In conjunction with ET observations of comparable mass IMBH-IMBH mergers, IMRIs could shed light on the hierarchical growth of structure, in particular to distinguish between light and heavy seed models, as described in more detail at the end of Section~\ref{s2.3}. 

A further potentially exciting payoff of ET IMRI observations will be to test whether the central object in an IMRI is described by the Kerr metric of general relativity. During the inspiral, the CO traces out the geometry of the spacetime of the central object. Hence, the GWs emitted during the inspiral encode a map of the spacetime. In~\cite{geroch,hansen} it was shown that any stationary, axisymmetric, vacuum spacetime in relativity can be decomposed into mass, \(M_{\ell}\), and current, \(S_{\ell}\), multipole moments. Ryan \cite{ryan} went on to show that, for nearly equatorial, nearly circular inspirals, these multipole moments are redundantly encoded in gravitational wave observables, namely the periapsis precession frequency, the orbital plane precession frequency and the gravitational wave energy spectrum. The multipole moments of a Kerr black hole are completely determined by its mass, \(M\), and spin, \(S_1\), through the relation \cite{hansen}
\begin{equation}
  M_{\ell}+\textrm{i}S_{\ell}=M\left(\textrm{i}a\right)^{\ell},
\label{mul}
\end{equation}
\noindent where  \(a=S_1/M\) is the reduced spin of the black hole. Extracting three moments of the spacetime from GW emission, and finding them inconsistent with \eqref{mul}, would suffice to demonstrate that the central object is not a Kerr black hole. Tests of this nature have been considered for observations of EMRI systems with LISA. Using a kludge model that included a non--Kerr value for the quadrupole moment of the central black hole, Barack and Cutler \cite{nonkerr} showed that LISA could measure the quadrupole moment, \(Q=-S^2/M\), of the central black hole to an accuracy \(\Delta Q/M^3\sim 10^{-3}\), while simultaneously measuring the mass and spin to an accuracy \(\sim 10^{-4}\). In the context of IMRIs detected by Advanced LIGO, it has been shown that it is possible to measure an \(0(1)\) fractional deviation in the mass quadrupole moment for typical systems \cite{brown}. ET could improve this precision by more than an order of magnitude, since ET's enhanced sensitivity will allow a significantly larger number of gravitational--wave cycles to be observed than will be possible with Advanced LIGO. For instance, for a \(1M_{\odot} + 100M_{\odot}\) system Advanced LIGO could measure \(\sim 500\) cycles until plunge. In contrast, ET with a low-frequency cut-off at $1$Hz will observed up to \(\sim 25000\) cycles\cite{finnthorne}. Further work is required to properly quantify the ability of ET to carry out these sorts of tests on the nature of IMBHs, and this should be pursued in the future.

Our results are a first attempt to explore the precision of IMRI parameter estimation that will be achievable with the Einstein Telescope. Our results have been derived using a particular waveform model, which reflects the best of what is currently available, and combines results from both comparable mass binaries and extreme-mass-ratio inspirals. These waveforms are unlikely to be accurate enough to be used in a search to recover source parameters, but they should capture most of the main features of true IMRI waveforms and therefore provide a decent estimate of the level of precision that could be achieved by GW measurements. The waveform models could be improved in various ways, by including conservative corrections, by generalizing the waveform models to consider both eccentric orbits and orbits inclined to the equatorial plane, and by including the leading order effects of the spin of the smaller object. The EOB model can also now be extended to spinning systems, for circular-equatorial inspirals at least~\cite{nic}, which will provide further important consistency checks for the transition model. These will be important improvements to consider in the future in order to confirm the present results, and to extend the calculations to generic IMRI systems.

\section*{Acknowledgments}
E.H. is supported by CONACyT. J.G.'s work is supported by the Royal Society.

\bibliography{paper2c}
\clearpage
\appendix

\section{Dependence of parameter estimation errors on network configuration}
This Appendix contains  Tables~\ref{conf2}--\ref{conf4}. These Tables show how the expected parameter measurement errors depend on the network configuration. We explore how parameter estimation accuracies are modified for five network configurations,  C1--C5.  These configurations are, C1: one ET at the geographic location of Virgo; C2: as configuration C1 plus a right--angle detector at the location of LIGO Livingston;  C3: as configuration C1 plus another ET at the location of LIGO Livingston;  and C4: as configuration C2 plus another right--angle detector in Perth. We denote the reference 3-ET network as configuration C5.

\begin{table}[thb]
\begin{tabular}{|c|c|c|c|c|c|c|c|c|c|c|c|}
\hline\multicolumn{2}{|c|}{}&\multicolumn{10}{c|}{Statistics of distribution for error \(\Delta X\) in parameter \(X=\)}\\\cline{3-12}
\multicolumn{2}{|c|}{Model}&$\ln(m)$&$\ln(M)$&$q$&$t_0$&$\phi_0$&$\theta_S$&$\phi_S$&$\theta_K$&$\phi_K$&$\ln(D)$\\\hline
&Mean      &1.26e-4     &7.76e-5        &1.96e-4        &2.63e-2        &2.01e-1        &3.7418        &4.1009        &6.6255        &7.4131        &9.2443\\\cline{2-12}
&St. Dev.  &6.54e-5     &4.57e-5        &9.27e-5        &1.30e-2        &1.09e-1        &4.0925        &4.7919        &4.5011        &5.2979        &6.6406\\\cline{2-12}
C1&L.Qt.   &6.73e-5     &4.47e-5        &1.84e-4        &1.73e-2        &1.29e-1        &0.5762        &0.4917        &2.8422        &3.5419        &5.2586\\\cline{2-12}
&Med.      &9.93e-5     &7.08e-5        &2.19e-4        &2.51e-2        &1.85e-1        &2.0259        &2.0586        &5.8983        &6.9206        &8.5296\\\cline{2-12}
&U. Qt.    &1.66e-4     &1.17e-4        &2.36e-4        &3.54e-2        &2.56e-1        &6.1471        &6.1689        &8.9838        &11.4924       &12.6824\\\hline

&Mean      &1.09e-4     &6.61e-5        &1.84e-4        &1.77e-2        &1.64e-1        &5.85e-2        &7.58e-2        &1.70e-1        &2.34e-1        &1.98e-1\\\cline{2-12}
&St. Dev.  &5.84e-5     &3.73e-5        &9.13e-5        &1.02e-2        &8.48e-2        &3.08e-2        &4.71e-2        &1.40e-1        &2.78e-1        &1.53e-1\\\cline{2-12}
C2&L.Qt.   &6.17e-5     &4.07e-5        &1.44e-4        &1.34e-2        &1.14e-1        &3.30e-2        &3.56e-2        &6.64e-2        &7.24e-2        &8.31e-2\\\cline{2-12}
&Med.      &8.81e-5     &6.17e-5        &1.90e-4        &1.65e-2        &1.50e-1        &5.22e-2        &6.02e-2        &1.28e-1        &1.81e-1        &1.45e-1\\\cline{2-12}
&U. Qt.    &1.46e-4     &1.02e-4        &2.19e-4        &2.33e-2        &2.17e-1        &7.73e-2        &1.04e-1        &2.36e-1        &5.01e-1        &2.50e-1\\\hline

&Mean      &9.82e-5     &6.31e-5        &1.81e-4        &1.65e-2        &1.50e-1        &4.36e-2        &4.95e-2        &8.71e-2        &1.44e-1        &1.05e-1\\\cline{2-12}
&St. Dev.  &5.28e-5     &3.48e-5        &9.09e-5        &9.11e-3        &7.71e-2        &1.55e-2        &2.15e-2        &6.34e-2        &1.39e-1        &7.11e-2\\\cline{2-12}
C3&L.Qt.   &5.62e-5     &3.80e-5        &1.29e-4        &1.28e-2        &1.17e-1        &2.95e-2        &3.06e-2        &4.36e-2        &5.37e-2        &5.46e-2\\\cline{2-12}
&Med.      &8.11e-5     &5.62e-5        &1.86e-4        &1.54e-2        &1.47e-1        &4.57e-2        &4.49e-2        &6.30e-2        &1.28e-1        &7.94e-2\\\cline{2-12}
&U. Qt.    &1.31e-4     &9.77e-5        &2.14e-4        &2.13e-2        &1.95e-1        &5.72e-2        &6.54 e-2       &1.14e-1        &2.69e-1        &1.31e-1\\\hline

&Mean      &9.79e-5     &6.03e-5        &1.81e-4        &1.65e-2        &1.50e-1        &3.97e-2        &4.67e-2        &8.74e-2        &1.41e-1        &1.02e-1\\\cline{2-12}
&St. Dev.  &4.78e-5     &3.37e-5        &8.81e-5        &8.81e-3        &5.84e-2        &1.18e-2        &1.72e-2        &6.33e-2        &1.20e-1        &6.84e-2\\\cline{2-12}
C4&L.Qt.   &5.66e-5     &3.74e-5        &1.29e-4        &1.31e-2        &1.17e-1        &3.03e-2        &3.17e-2        &4.45e-2        &4.89e-2        &5.24e-2\\\cline{2-12}
&Med.      &7.61e-5     &5.83e-5        &1.82e-4        &1.69e-2        &1.50e-1        &3.96e-2        &4.41e-2        &6.44e-2        &1.00e-1        &8.12e-2\\\cline{2-12}
&U. Qt.    &1.20e-4     &9.35e-5        &2.08e-4        &2.13e-2        &1.95e-1        &4.97e-2        &5.89e-2        &1.17e-1        &1.96e-1        &1.34e-1\\\hline

&Mean      &9.41e-5     &5.99e-5        &1.80e-4        &1.65e-2        &1.36e-1        &3.97e-2        &4.63e-2        &7.91e-2        &9.15e-2        &9.96e-2\\\cline{2-12}
&St. Dev.  &4.24e-5     &3.13e-5        &6.86e-5        &8.11e-3        &6.83e-2        &1.25e-2        &1.84e-2        &5.48e-2        &6.27e-2        &5.84e-2\\\cline{2-12}
C5&L.Qt.   &5.68e-5     &3.78e-5        &1.31e-4        &1.28e-2        &1.05e-1        &3.05e-2        &3.18e-2        &3.86e-2        &4.91e-2        &5.11e-2\\\cline{2-12}
&Med.      &8.96e-5     &5.85e-5        &1.78e-4        &1.69e-2        &1.37e-1        &3.96e-2        &4.59e-2        &5.74e-2        &7.81e-2        &7.24e-2\\\cline{2-12}  
&U. Qt.    &1.15e-4     &9.01e-5        &2.07e-4        &2.08e-2        &1.67e-1        &5.17e-2        &5.98e-2        &1.04e-1        &9.95e-2        &1.20e-1\\\hline
\end{tabular}
\caption{Summary of Monte Carlo results for parameter estimation errors.  We show the mean, standard deviation, median and quartiles of the distribution of the error in each parameter.   Results are given for a \(m=1.4M_{\odot}\)  CO inspiralling into a \(M=100M_{\odot}\) IMBH   with  spin parameter \(q=0.3\), and assuming four alternative configurations for the detector network, C1--C4, as described above. Configuration C5 is the network of three ETs which has been used for all results elsewhere in this paper.}
\label{conf2}
\end{table}

\begin{table}[thb]
\begin{tabular}{|c|c|c|c|c|c|c|c|c|c|c|c|}
\hline\multicolumn{2}{|c|}{}&\multicolumn{10}{c|}{Statistics of distribution for error \(\Delta X\) in parameter \(X=\)}\\\cline{3-12}
\multicolumn{2}{|c|}{Model}&$\ln(m)$&$\ln(M)$&$q$&$t_0$&$\phi_0$&$\theta_S$&$\phi_S$&$\theta_K$&$\phi_K$&$\ln(D)$\\\hline
&Mean      &2.14e-3     &5.01e-4        &2.34e-3        &2.08e-2        &2.34e-1        &1.5332        &2.3089        &2.3442        &2.7840        &2.5105\\\cline{2-12}
&St. Dev.  &1.07e-3     &3.51e-4        &1.03e-3        &7.33e-3        &1.30e-1        &2.0017        &3.1318        &1.9584        &2.2266        &1.8864\\\cline{2-12}
C1&L.Qt.   &1.78e-3     &2.63e-4        &1.95e-3        &1.02e-2        &1.62e-1        &0.1341        &0.2344        &0.8615        &1.1489        &1.1982\\\cline{2-12}
&Med.      &2.51e-3     &5.01e-4        &2.34e-3        &1.62e-2        &2.45e-1        &0.6232        &0.8482        &1.9277        &2.0915        &1.8508\\\cline{2-12}
&U. Qt.    &3.16e-3     &8.51e-4        &2.63e-3        &2.69e-2        &3.38e-1        &2.1273        &3.1569        &3.4303        &3.8956        &3.2680\\\hline

&Mean      &1.70e-3     &4.17e-4        &2.19e-3        &1.31e-2        &1.51e-1        &6.08e-2        &9.23e-2        &1.99e-1        &2.28e-1        &2.39e-1\\\cline{2-12}
&St. Dev.  &1.15e-3     &3.13e-4        &7.79e-4        &6.46e-3        &7.06e-2        &3.02e-2        &5.75e-2        &1.73e-1        &1.90e-1        &1.86e-1\\\cline{2-12}
C2&L.Qt.   &1.26e-3     &2.09e-4        &1.82e-3        &8.71e-3        &1.12e-1        &3.50e-2        &4.27e-2        &8.03e-2        &8.81e-2        &9.77e-2\\\cline{2-12}
&Med.      &2.29e-3     &3.72e-4        &2.19e-3        &1.28e-2        &1.47e-1        &5.95e-2        &7.01e-2        &1.29e-1        &1.63e-1        &1.63e-1\\\cline{2-12}
&U. Qt.    &2.88e-3     &7.24e-4        &2.51e-3        &2.04e-2        &2.08e-1        &7.81e-2        &1.37e-1        &2.51e-1        &2.87e-1        &3.16e-1\\\hline

&Mean      &1.48e-3     &3.80e-4        &2.09e-3        &1.23e-2        &1.34e-1        &4.38e-2        &5.56e-2        &1.11e-1        &1.31e-1        &1.32e-1\\\cline{2-12}
&St. Dev.  &9.07e-4     &2.43e-4        &7.01e-4        &5.76e-3        &4.70e-2        &1.56e-2        &2.52e-2        &9.34e-2        &1.08e-1        &1.02e-1\\\cline{2-12}
C3&L.Qt.   &1.05e-3     &2.14e-4        &1.78e-3        &8.13e-3        &1.04e-1        &2.99e-2        &3.41e-2        &5.37e-2        &5.60e-2        &6.74e-2\\\cline{2-12}
&Med.      &2.14e-3     &3.55e-4        &2.14e-3        &1.25e-2        &1.28e-1        &4.18e-2        &4.95e-2        &7.82e-2        &9.88e-2        &9.70e-2\\\cline{2-12}
&U. Qt.    &2.69e-3     &6.76e-4        &2.45e-3        &1.90e-2        &1.73e-1        &5.71e-2        &7.43e-2        &1.35e-1        &1.70e-1        &1.76e-1\\\hline

&Mean      &1.51e-3     &3.72e-4        &2.09e-3        &1.20e-2        &1.28e-1        &3.97e-2        &5.31e-2        &1.05e-1        &1.21e-1        &1.20e-1\\\cline{2-12}
&St. Dev.  &8.06e-4     &2.51e-4        &6.08e-4        &5.09e-3        &4.75e-2        &1.13e-2        &1.96e-2        &8.40e-2        &9.73e-2        &9.51e-2\\\cline{2-12}
C4&L.Qt.   &1.15e-3     &1.95e-4        &1.70e-3        &7.59e-3        &1.04e-1        &3.16e-2        &3.49e-2        &5.24e-2        &5.71e-2        &6.66e-2\\\cline{2-12}
&Med.      &2.09e-3     &3.39e-4        &2.14e-3        &1.23e-2        &1.31e-1        &4.06e-2        &4.85e-2        &8.21e-2        &9.40e-2        &9.82e-2\\\cline{2-12}
&U. Qt.    &2.63e-3     &6.56e-4        &2.45e-3        &1.86e-2        &1.62e-1        &4.99e-2        &7.01e-2        &1.39e-1        &1.59e-1        &1.72e-1\\\hline

&Mean       &1.41e-3     &3.63e-4       &2.09e-3        &1.19e-2        &1.05e-1        &3.95e-2        &5.12e-2        &9.57e-2        &1.08e-1        &1.12e-1\\\cline{2-12}
&St. Dev.   &6.39e-4     &1.87e-4       &5.72e-4        &5.07e-3        &5.20e-2        &1.19e-2        &2.03e-2        &6.74e-2        &8.75e-2        &7.67e-2\\\cline{2-12}
C5&L.Qt.    &1.07e-3     &2.00e-4       &1.65e-3        &7.41e-3        &7.83e-2        &2.96e-2        &3.43e-2        &5.74e-2        &5.75e-2        &5.93e-2\\\cline{2-12}
&Med.       &1.58e-3     &3.47e-4       &2.13e-3        &1.24e-2        &9.38e-2        &3.86e-2        &4.79e-2        &6.91e-2        &8.79e-2        &8.82e-2\\\cline{2-12} 
&U. Qt.     &1.74e-3     &6.03e-4       &2.40e-3        &1.41e-2        &1.22e-1        &4.92e-2        &6.51e-2        &1.14e-1        &1.24e-1        &1.56e-1\\\hline
\end{tabular}
\caption{As Table~\ref{conf2}, but for binary systems with a CO of mass \(m=10M_{\odot}\), and a central IMBH of mass \(M=500M_{\odot}\).  }
\label{conf3}
\end{table}

\begin{table}[thb]
\begin{tabular}{|c|c|c|c|c|c|c|c|c|c|c|c|}
\hline\multicolumn{2}{|c|}{}&\multicolumn{10}{c|}{Statistics of distribution for error \(\Delta X\) in parameter \(X=\)}\\\cline{3-12}
\multicolumn{2}{|c|}{Model}&$\ln(m)$&$\ln(M)$&$q$&$t_0$&$\phi_0$&$\theta_S$&$\phi_S$&$\theta_K$&$\phi_K$&$\ln(D)$\\\hline
&Mean      &3.90e-4     &6.97e-5        &2.14e-4        &2.75e-2       &3.37e-1        &1.0248        &1.4132        &1.73984        &1.98796        &1.96526\\\cline{2-12}
&St. Dev.  &9.60e-5     &1.30e-5        &5.03e-5        &1.12e-2       &2.11e-1        &1.0433        &1.5308        &0.98158        &1.25075        &1.21136\\\cline{2-12}
C1&L.Qt.   &3.56e-4     &6.24e-5        &1.87e-4        &2.23e-2       &1.33e-1        &0.1554        &0.2162        &0.94480        &0.93815        &0.97202\\\cline{2-12}
&Med.      &4.04e-4     &7.02e-5        &2.12e-4        &2.69e-2       &2.76e-1        &0.6771        &0.8102        &1.60984        &1.78872        &1.76174\\\cline{2-12}
&U. Qt.    &4.43e-4     &7.79e-5        &2.32e-4        &3.80e-2       &6.02e-1        &1.5840        &2.2538        &2.42813        &2.82516        &2.74045\\\hline

&Mean      &3.37e-4     &6.92e-5        &2.10e-4        &1.81e-2       &1.38e-1        &5.69e-2        &8.02e-2        &1.93e-1        &2.37e-1        &1.88e-1\\\cline{2-12}
&St. Dev.  &1.31e-4     &1.29e-5        &5.00e-5        &9.60e-3       &1.07e-1        &2.85e-2        &5.13e-2        &1.61e-1        &2.05e-1        &1.31e-1\\\cline{2-12}
C2&L.Qt.   &2.41e-4     &6.23e-5        &1.91e-4        &1.62e-2       &6.26e-2        &3.24e-2        &3.67e-2        &7.81e-2        &9.54e-2        &8.84e-2\\\cline{2-12}
&Med.      &3.89e-4     &7.02e-5        &2.11e-4        &1.99e-2       &8.27e-2        &5.02e-2        &6.51e-2        &1.39e-1        &1.54e-1        &1.43e-1\\\cline{2-12}
&U. Qt.    &4.23e-4     &7.46e-5        &2.25e-4        &2.57e-2       &1.81e-1        &7.23e-2        &1.15e-1        &2.56e-1        &3.16e-1        &2.56e-1\\\hline

&Mean      &3.23e-4     &6.90e-5        &2.05e-4        &1.62e-2       &1.25e-1        &4.36e-2        &5.96e-2        &1.00e-1        &1.38e-1        &9.97e-2\\\cline{2-12}
&St. Dev.  &1.30e-4     &1.18e-5        &4.57e-5        &6.12e-3       &1.06e-1        &1.77e-2        &3.07e-2        &8.18e-2        &1.11e-1        &6.14e-2\\\cline{2-12}
C3&L.Qt.   &2.10e-4     &6.23e-5        &1.95e-4        &1.47e-2       &6.30e-2        &2.75e-2        &3.31e-2        &4.84e-2        &6.45e-2        &5.62e-2\\\cline{2-12}
&Med.      &3.75e-4     &6.97e-5        &2.06e-4        &1.77e-2       &8.12e-2        &3.96e-2        &5.75e-2        &6.82e-2        &9.11e-2        &7.89e-2\\\cline{2-12}
&U. Qt.    &4.13e-4     &7.58e-5        &2.19e-4        &2.23e-2       &2.45e-1        &5.82e-2        &7.43e-2        &1.29e-1        &1.73e-1        &1.39e-1\\\hline

&Mean      &3.21e-4     &6.85e-5        &2.03e-4        &1.20e-2       &1.23e-1        &4.06e-2        &5.88e-2        &9.12e-2        &1.24e-1        &9.74e-2\\\cline{2-12}
&St. Dev.  &1.19e-4     &1.01e-5        &4.43e-5        &5.33e-3       &7.95e-2        &1.47e-2        &2.81e-2        &6.56e-2        &1.03e-1        &6.36e-2\\\cline{2-12}
C4&L.Qt.   &2.30e-4     &6.25e-5        &1.93e-4        &1.07e-2       &5.88e-2        &2.99e-2        &3.46e-2        &4.64e-2        &6.38e-2        &5.09e-2\\\cline{2-12}
&Med.      &3.73e-4     &6.86e-5        &2.04e-4        &1.23e-2       &8.12e-2        &3.96e-2        &5.62e-2        &6.81e-2        &8.71e-2        &7.63e-2\\\cline{2-12}
&U. Qt.    &4.13e-4     &7.45e-5        &2.15e-4        &1.73e-2       &2.51e-1        &5.17e-2        &6.61e-2        &1.19e-1        &1.66e-1        &1.29e-1\\\hline

&Mean      &3.17e-4     &6.82e-5        &2.03e-4        &1.14e-2        &9.02e-2        &4.02e-2        &5.75e-2        &8.59e-2        &1.10e-1       &9.68e-2\\\cline{2-12} 
&St. Dev.  &1.03e-4     &8.51e-6        &4.31e-5        &5.22e-3        &5.32e-2        &1.28e-2        &2.84e-2        &6.22e-2        &9.12e-2       &6.10e-2\\\cline{2-12} 
C5&L.Qt.   &2.21e-4     &6.28e-5        &1.91e-4        &1.02e-2        &7.02e-2        &3.02e-2        &3.23e-2        &4.48e-2        &5.25e-2       &4.91e-2\\\cline{2-12} 
&Med.      &3.51e-4     &6.80e-5        &2.01e-4        &1.17e-2        &9.34e-2        &4.16e-2        &5.17e-2        &6.14e-2        &8.22e-2       &7.39e-2\\\cline{2-12}
&U. Qt.    &4.06e-4     &7.31e-5        &2.11e-4        &1.65e-2        &1.24e-1        &5.09e-2        &6.97e-2        &1.06e-1        &1.50e-1       &1.21e-1\\\hline
\end{tabular}
\caption{As Table~\ref{conf2}, but for binary systems with a central IMBH of mass \(M=500M_{\odot}\).  }
\label{conf4}
\end{table}

\end{document}